\documentclass{aa}  

\usepackage{graphicx}
\usepackage[varg]{txfonts}
\usepackage{natbib}

\usepackage{amsmath,array,amsfonts,mathrsfs}
\usepackage{float}
\usepackage{amssymb}
\usepackage{txfonts}
\usepackage{multirow}
\usepackage{pifont}
\usepackage{rotating}

\usepackage[normalem]{ulem}

\bibpunct{(}{)}{;}{a}{}{,}
\begin{document}
   \title{The XMM-Newton deep survey in the CDF-S VI.}

   \subtitle{Obscured AGN selected as infrared power-law galaxies}

   \author{N. Castell\'o-Mor
          \inst{1}
          \and
          F.J. Carrera\inst{1}
	\and 
	A.~Alonso-Herrero\inst{1}\fnmsep\thanks{Augusto G. Linares Senior Research Fellow}
	\and
	S.~Mateos\inst{1}
	\and
	X.~Barcons\inst{1}
	\and
    P.~Ranalli\inst{2,3}
    \and
    P.G.~P\'erez-González\inst{4}\fnmsep\thanks{Associate Astronomer at Steward Observatory, The University of Arizona, Tucson, AZ, USA}
	\and 
    A.~Comastri\inst{3}
    \and
    C.~Vignali\inst{5}
    \and
    I.~Georgantopoulos\inst{2}
    }
    
    \offprints{N. Castell\'o-Mor}

    \institute{Instituto de F\'\i sica de Cantabria (CSIC-UC), 39005 Santander, Spain
        \and
    Institute of Astronomy, Astrophysics, Space Applications and Remote Sensing, National Observatory of Athens, Palaia Penteli, 15236, Athens, Greece
        \and
    INAF - Osservatorio Astronomico di Bologna, Via Ranzani 1, Bologna, 40127, Italy
        \and
	Departamento de Astrof\'\i sica, Facultad de CC. F\'\i sicas, Universidad Complutense de Madrid, 28040 Madrid, Spain
        \and
    Physics \& Astronomy Department, University of Bologna, Viale Berti Pichat 6/2, Bologna, 40127, Italy
             }

   \date{Received 2012; accepted xxxx}

 
  \abstract
  { Accretion onto supermassive black
  holes is believed to occur mostly in obscured Active Galactic Nuclei
  (AGN).  Such objects are proving rather elusive in surveys of
  distant galaxies, including those at X-ray energies.  }
  {Our main goal is to determine whether the revised IRAC criteria of
    \citet{Donley2012} (objects with an infrared (IR) power-law
    spectral shape), are effective at selecting X-ray type-2 AGN
    (i.e., absorbed $N_H>10^{22}$ cm$^{-2}$).  }
  { We present the results from the X-ray spectral analysis of 147 AGN
   selected by cross-correlating the highest spectral quality
   ultra-deep {\it XMM-Newton} and the {\it Spitzer}/IRAC catalogues
   in the {\it Chandra} Deep Field South. Consequently it is
     biased towards sources with high S/N X-ray spectra.  In order to
   measure the amount of intrinsic absorption in these sources, we
   adopt a simple X-ray spectral model that includes a power-law
   modified by intrinsic absorption at the redshift of each source and
   a possible soft X-ray component. }
 {We find 21/147 sources to be heavily absorbed but the uncertainties
   in their obscuring column densities do not allow us to confirm
   their Compton-Thick nature without resorting to additional
   criteria. Although IR power-law galaxies are less numerous in our
   sample than IR non-power-law galaxies (60 versus 87 respectively),
   we find that the fraction of absorbed ($N_{H}^{intr}> 10^{22}\,
   {\rm cm}^{-2}$) AGN is significantly higher (at about 3 sigma
   level) for IR-power-law sources ($\sim2/3$) than for those sources
   that do not meet this IR selection criteria ($\sim1/2$).  This
   behaviour is particularly notable at low luminosities, but it
   appears to be present, although with a marginal significance, at
   all luminosities.  }
{ We
therefore conclude that the IR power-law method is efficient in
finding X-ray-absorbed sources. We would then expect that the
long-sought dominant population of absorbed AGN is abundant among
IR power-law spectral shape sources not detected in X-rays.}

   \keywords{X-rays: galaxies - Galaxies: active - Surveys - infrared: galaxies}

   \maketitle
%
\newcommand{\xmm}{{\itshape XMM-Newton}}
\newcommand{\irac}{IRAC}
\newcommand{\spitzer}{{\itshape Sptizer} IRAC}
\newcommand{\chandra}{{\itshape Chandra}}
\newcommand{\xspec}{{\itshape  XSPEC}}
\newcommand{\phabs}{\textrm{PHABS}}
\newcommand{\powerlaw}{\textrm{POWERLAW}}
\newcommand{\zbbody}{\textrm{ZBBODY}}
\newcommand{\zphabs}{\textrm{ZPHABS}}
\newcommand{\unh}{${\rm cm}^{-2}$}

\section{Introduction}
\label{section1}

Obscured active galactic nuclei (AGN) have attracted much attention
for a few decades, partly because a significant fraction of the power
from accretion onto black holes is believed to be shrouded by
circumnuclear matter (e.g. \citealt{Fabian1999}). Such a population of
obscured AGN is expected to be a major contributor to the cosmic X-ray
background (XRB), \citep{Gilli2007}. Since X-rays come from very close
to the central engine in AGN and X-ray emission is not severely
contaminated by the host galaxy, X-ray surveys are often used as an
efficient method for identifying reliable and fairly complete samples
of AGN \citep{Mushotzky2004,Brandt2005}.  Large numbers of AGN are
found in deep X-ray surveys, where the observed AGN sky density is
about an order of magnitude higher than that found at any other
wavelength (e.g. \citealt{Steidel2002, Bauer2004}). These X-ray point
sources can account for most of the observed XRB intensity at energies
below 8~keV.  X-ray surveys with \chandra~ and \xmm~ have now resolved
the totality of the XRB below $\sim 2\, {\rm keV}$
(e.g. \citealt{Bauer2004,Mushotzky2004,Hickox2006}). This fraction,
however, drops to $\sim 50$\% at photon energies above $8\, {\rm keV}$
(\citealt{Worsley2004,Worsley2005}). The majority of the X-ray point
sources detected in the $2$-$8$ keV band are moderately obscured AGN
($N_{H} \lesssim 3\times 10^{23}$ cm$^{-2}$; e.g.
\citealt{Szololy2004,Barger2005,Tozzi2006,Mateos2005}). On the other
hand, an additional population of heavily obscured ($N_{H}\gtrsim
3\times 10^{23}$ cm$^{-2}$), or even Compton-Thick ($N_{H} > 1.5
\times 10^{24}$ cm$^{-2}$) AGN at cosmological distances, which are
missed by conventional quasar surveys, is required by AGN synthesis
models of the XRB
(e.g. \citealt{Gilli2007,Treister2009,Luo2011}). These are the best
candidates to fill the residual 50\% of the XRB above $8$ keV, not
accounted for by source populations in existing X-ray One fundamental
ingredient in our understanding of the AGN population is the ratio of
obscured to unobscured AGN, and how this ratio depends on parameters
like intrinsic luminosity or redshift. Thus, a measurement of the
fraction of obscured AGN and its possible dependence on critical
parameter's can be used to study AGN structure and to probe the
connection between AGN activity and formation of the host galaxy.  The
number density of these heavily obscured and Compton-Thick AGN is
expected to outnumber unobscured AGN by a ratio of $\sim 2$-$4:1$
(e.g. \citealt{Maiolino1995,Comastri2001_XRB,Gilli2001,Gilli2007,Xue2012}),
though the exact value is still not well constrained. Deep X-ray
observations can only detect the most luminous of these hard X-ray
sources, and even the deepest X-ray surveys likely miss large
populations of heavily obscured AGN
\citep{Tozzi2006,Georgantopoulos2007,Georgantopoulos2009,Comastri2011,Feruglio2011,DellaCeca2008}.
Thus, a significant fraction of the AGN population probably remains
undetected even in X-rays.\\

According to the Unified AGN model, AGN are thought to have a dusty
environment surrounding an optically and X-ray-bright accretion disc
around a super-massive black hole ($M_{BH}\gtrsim 10^6\,M_{\odot}$,
e.g.  \citealt{Antonucci1993,Urry1995}).  This dusty material is
expected to be distributed in a torus-shaped structure centred in the
super-massive black hole.  The torus provides anisotropic obscuration
of the central region so that sources viewed through a gas and
dust-free line of sight are recognized as type-1 AGN, and those
obscured by dust are classified as type-2 AGN, based on the
characteristics of their optical spectra.  Incident X-ray, optical and
UV radiation from the central engine can be absorbed and reprocessed
by the circumnuclear dust.  An obscuring dusty torus should re-radiate
in the infrared (IR) a significant fraction of the nuclear luminosity
it absorbs, and the continua from most AGN indeed show significant IR
emission \citep{Barvainis1987,Granato1994,Nenkova2002}.  Galaxies
dominated by AGN emission typically show a power-law ($f_{\nu}\propto
\nu^{\alpha}$, where $\alpha$ is the spectral index) spectral energy
distribution (SED) in those ranges, although with a variety of slopes
(from $\alpha=-0.5$ up to $\alpha=-3$). While the power-law locus
itself extends to bluer slopes, luminous AGN are expected to display
red slopes of $\alpha\leq-0.5$ \citep{AlonsoHerrero2006,Donley2007}.
Of course, this does not necessarily mean that the spectral index is
the same in the near- and mid-IR.  It has been recognized that the
ultraviolet (UV) and optical continuum of luminous QSOs could be
described with a power law that can continue all the way into the
near-, mid-, and even far-IR (see, e.g.,
\citealt{Neugebauer1979,Elvis1994}).  The power-law-like emission at
short wavelengths ($0.1 \lesssim \lambda \lesssim 1 \mu$m), which is an
upturn towards the big blue bump tail comes from the AGN accretion
disc.  A broad continuum excess is detected in the near-IR continuum
of galaxies in the rest frame at $2-5\,\mu$m
\citep{Lu2003,Helou2004,Magnelli2008}, which is attributed to radiation
reprocessed by the dusty torus. Clumpy torus models
\citep{Nenkova2008} predict nearly isotropic emission at IR
wavelengths $\geq 12\mu$m for any torus model parameters.\\

Thus, the IR spectral range provides a powerful, complementary method
for identifying AGN over a wide range of intrinsic obscuration that
might not be detected in the X-ray band.  Various IR selections have
been employed extensively to search for AGN in which strong
re-radiation from obscuring dust is expected
\citep{Rieke1981,Edelson1986,Lacy2004,Stern2005,Barvainis1987,PerezGarcia1998,AlonsoHerrero2006,Mateos2012,Stern2012,Assef2012,Yan2013}.
With the \emph{Spitzer} mission \citep{Werner2004}, deep mid-IR data
in multiple bands have been obtained in various cosmological survey
fields with the IR Array Camera (IRAC, \citealt{Fazio2004}) and
Multiband Imaging Photometer (MIPS, \citealt{Rieke2004}) instruments.
The \spitzer~power-law selection
\citep{AlonsoHerrero2006,Donley2007,Donley2012} chooses sources whose
IRAC SEDs follow a power law over a wide range of slopes ($\alpha$).
This is because, galaxies dominated by AGN emission typically exhibit
a characteristic red power-law SED across the IRAC bands due to the
superposition of blackbody emission from the AGN-heated dust.  A
problem commonly encountered when studying AGN properties based on IR
observations is the significant contribution of the host galaxy to the
near- and mid-IR
\citep{Kotilainen1992,AlonsoHerrero1996,Franceschini2005}. However, in
high-luminosity objects where the AGN outshines the host galaxy by a
large factor in the rest-frame optical and near- and mid-IR, this
should not be the case.\\

In this paper, we attempt to find midly/heavily obscured AGN in the
ultra-deep \xmm~ observations (carried out at $2$-$10$~keV) in the
\chandra~ Deep Field South (CDF-S) based on an IR-colour
selection. For the mid-IR, we use the deep data taken with {\itshape
  Spitzer}/IRAC at $3.6$-$8.0$~$\mu$m in this field and catalogued by
\citet{PerezGonzalez2008}.  We also compare the X-ray properties of
the sources detected both in X-rays and in the four IRAC bands with
and without a power-law-like continuum shape.  The goal is to
  explore and quantify the efficiency in finding obscured ($N_{H}>
  10^{22}\, {\rm cm}^{-2}$) AGN within the X-ray-detected sources and
  especially whether it is effective at selecting type-2 AGN at some
  luminosity range when looking at X-ray emitting sources that display
  a power-law SED in the IR. Any discussion about the origin of
  the presumably more heavily obscured set of IR galaxies not detected
  in X-rays is beyond the scope of this paper.  There are other
papers in these series
\citep{Comastri2011,Iwasawa2012,Georgantopoulos2013} investigating the
heavily obscured and Compton-Thick nature of the sources in this
ultra-deep \xmm{} data, using different spectral models and additional
criteria, obtaining compatible results, as we will discuss below.\\

This paper is organized as follows. We describe the parent X-ray and
IR catalogues in Section 2, and we explain our method to select
sources with an IR power-law-shape \citep{Donley2012}. In Section 3,
we present our X-ray spectral analysis of the sources and the spectral
findings are discussed in Section4.  We discuss in more detail the
absorption column density distribution and the obscured AGN fraction
as a function of the intrinsic X-ray luminosity and the near- and
mid-IR-continuum shape, in Section 5. We summarize our conclusions in
Section 6. Throughout the paper errors are 90\% confidence for a
single parameter (i.e., $\Delta\chi^2=2.71$, \citealt{Avni1976}),
unless otherwise stated. We estimated the most probable value for the
fractions using a Bayesian approach and the binomial distribution from
\citet{Wall2008} for which the quoted errors are the narrowest
interval that includes the mode and encompasses 90\% of the
probability.  We assume the concordance cosmological model with
$H_0=70.5$ km s$^{-1}$ Mpc$^{-1}$, $\Omega_M=0.27$, and
$\Omega_{\Lambda} = 0.73$ \citep{Komatsu2009}.


\section{Sample}
\label{section2}

To study the X-ray properties of a sample of IR power-law AGN in the
CDF-S and the Extended CDF-S (E-CDF-S) fields using ultra-deep \xmm~
observations we took advantage of the deep {\itshape Spitzer} images
available in that region.  The CDF-S and the E-CDF-S areas were
surveyed with \xmm~ during different epochs spread over almost nine
years (see \citet{Ranalli2013} for more details).  The average
Galactic column density towards the CDF-S is $0.9\times10^{20}$
cm$^{-2}$ \citep{Dickey1990} providing a relatively clean vision of
the extragalactic X-ray sky even down to soft X-ray energies.

\subsection{Infrared}
\label{subsecInfrared}
The near- and mid-IR source catalogue used for this work has been
built from {\itshape Spitzer}/IRAC observations (the selection being
made at 3.6 $\mu$m and 4.5 $\mu$m). The total surveyed area is 664
arcmin$^2$. This IRAC sample is 75\% complete down to 1.6 $\mu$Jy. The
data are described in detail in \citet{PerezGonzalez2008}. A total of
23,044 non-flagged IRAC sources are selected from this sample with a
high signal-to-noise ratio (S/N$>5$) in each of the four bands (3.6,
4.5, 5.8, and 8.0 $\mu$m).

\subsection{X-rays}
\label{subsecXraysData}

The X-ray data presented in this paper were obtained from the CDF-S
\xmm~ survey.  An extended and detailed description of the full data
set, including the data analysis and reduction as well as the X-ray
catalogue will be published in \citet{Ranalli2013}. Briefly, the bulk
of the X-ray observations were made between July 2008 and March 2010,
and were combined with archival data taken between July 2001 and
January 2002.  The total net integration time (after removal of
background flares) is $\sim$2.8 Ms and $\sim$2.5 Ms for the EPIC MOS
and pn detectors, respectively. Standard \xmm~ software and procedures
were used for the analysis of the data. The source catalogue we use
for this paper contains X-ray sources detected in the 2-10 keV band
with conservative detection criteria: $>$8 $\sigma$ PWXDetect
significance and an exposure time $>$1Ms.  These requirements resulted
originally in 171 X-ray-detected sources. We then sub-selected the 150
objects for which spectral data are available for at least one of the
three EPIC cameras: pn, MOS1 or MOS2 (see \citealt{Comastri2013} for
more information about the definition of the spectral catalogue). 
  The exclusion of these 21 sources, spanning a wide range of redshift
  up to $z\sim$3, does not intruduce any additional bias.  Finally,
we excluded 3 further objects for which the redshift is unknown. Our
final X-ray sample contained then 147 AGN with spectroscopic and/or
photometric redshifts, all with $>$180 counts in the observed 0.5-9
keV energy band.  We note that the sample is not statistically
  complete as a result of the primary selection criteria, however, the
  aim of the paper is independent of any selection bias.\\

\begin{figure*}
    \centering
    \includegraphics[width=0.485\textwidth]{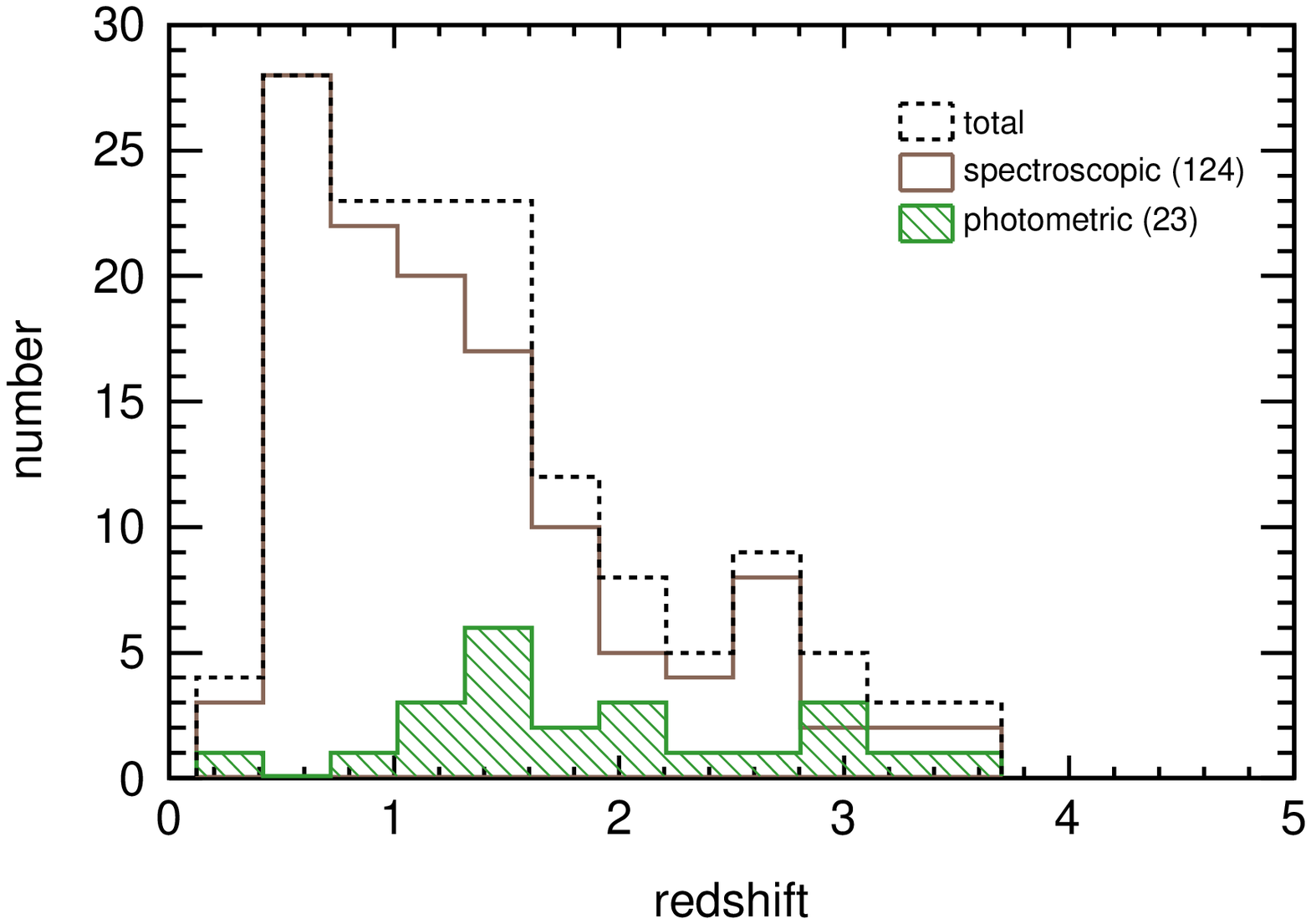}
    \includegraphics[width=0.485\textwidth]{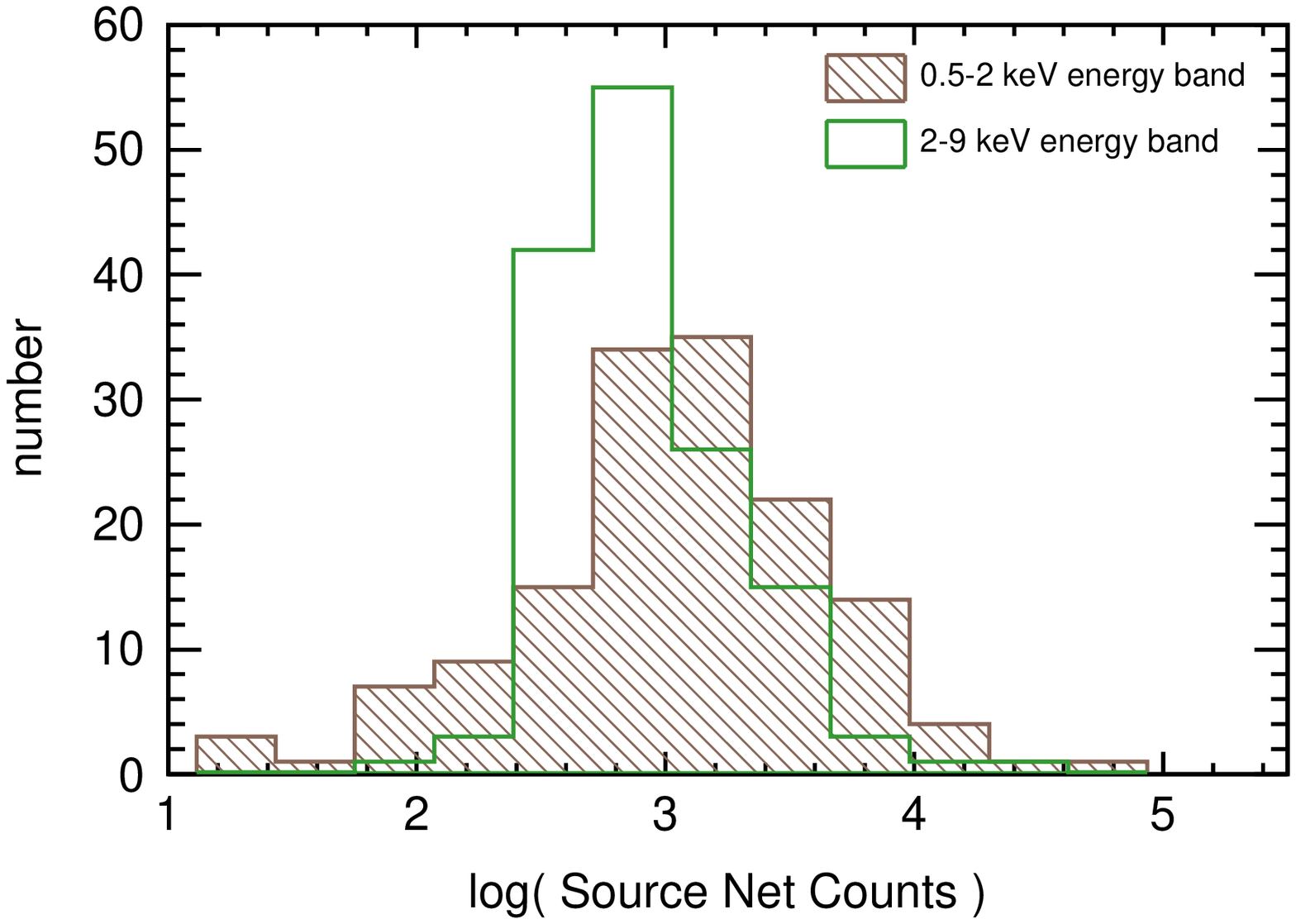}
    \caption{\emph{Left:} Spectroscopic and photometric redshift
      distribution as \emph{solid-line} and \emph{hatched} histograms,
      respectively. The dashed-line histogram represents the
      distribution of the redshift, regardless of whether it is
      photometric or spectroscopic. \emph{Right:} source net count
      distribution for both the 0.5-2 keV and 2-9 keV energy bands as
      \emph{hatched} and \emph{empty} histograms, respectively.}
    \label{fig:sample_properties}
\end{figure*}

Hereafter we use ``ID210= for the identification number of X-ray
sources listed in \citet{Ranalli2013}.  Spectroscopic redshifts are
available for $124$ objects while photometric redshifts were estimated
by various papers for 23 further objects. The photometric and
spectroscopic redshifts adopted in this paper are taken from
\citet{Ranalli2013}.  The redshift distribution of the sample is shown
in Figure \ref{fig:sample_properties}. The median redshift is
$1.40$. The distribution of the net (i.e., background subtracted)
source counts in both the 2-9 keV and 0.5-2 keV energy bands are also
given the same figure. \\

A classification purely based on optical properties has not been
possible for the majority of the sources. Correlations between optical
and X-ray properties of obscured and unobscured AGN show a very good
match ($\sim$80$\%$ or more) between optical reddening and X-ray
absorption (e.g., \citealt{Tozzi2006,Page2003,Perola2004}). Usually, relatively
low S/N unabsorbed AGN X-ray spectra are fit by simple power law
models. The effect of increasing column density is that the soft part
of the spectrum is more and more suppressed due to photometric
absorption, i.e., the spectrum becomes harder at high $N_{H}$.  It is
customary to define an X-ray unabsorbed AGN by $N_{H} < 10^{22}$
cm$^{-2}$, whereas absorbed AGN satisfy $N_{H} \geq 10^{22}$
cm$^{-2}$. This ``borderline'' value is a rather conservative upper
limit to the amount of absorbing gas across the interstellar medium of
a typical host galaxy's line of sight.

\subsection{Cross-correlation}

To identify X-ray/IR associations, we analysed the CDF-S ultra-deep
\xmm~ observations covering the sky positions of the IRAC
catalogue. We crosscorrelated both catalogues applying the method
developed by \citet{Pineau2011}.  The method is based on a likelihood
ratio (LR) technique. For a given source, this method provides the
probability of association for each candidate counterpart, which is a
function of the positional errors, relative distance and the local
density of potential counterparts. The much higher density of IRAC
sources with respect to that of \xmm~ sources and the large
uncertainties in the \xmm's positions make this cross-correlation
exercise particularly difficult. To facilitate our task, we
used the cross-correlation between CDF-S \xmm~ and the extended CDF-S
\chandra~ catalogue \citep{Ranalli2013} as a first step to find the
CDF-S \xmm~ counterparts to our IRAC targets. The reason for this is
that the positions of the \chandra~ sources are much more accurate (a
fraction of an arcsec) than those from \xmm~ (several arcsec,
\citealt{Lehmer2005,Xue2011}), while the densities of both \chandra~
and \xmm~ catalogues are similar. In practice this means that the
associations between \chandra~ and \xmm~ X-ray sources are highly
reliable and this step does not introduce any uncertainty in our
process to associate \xmm~ to IRAC sources.  We then use the {\itshape
  Chandra}'s positions instead of the less accurate \xmm~ positions to
search for IRAC counterparts.\\ {\itshape Spitzer}/IRAC data cover
very well over $95\%$ of the \xmm~ exposure area used in this
work. Only 6 X-ray sources fall in regions where there is partial or
no coverage by the {\itshape Spitzer}/IRAC catalogue data. The
catalogue of IR sources used is a compilation resulting from a number
of different {\itshape Spitzer}/IRAC surveys in the E-CDF-S area, of
different depths.  
We found one single IRAC counterpart for each X-ray source, and
therefore our final sample contains $147$ cross-correlations with a
single counterpart. Thus, all sources have an X-ray spectrum from the
CDF-S \xmm~ observations, {\itshape Spitzer}/IRAC fluxes in the 4
bands and a spectroscopic or photometric redshift.  The mean value
X-ray-\chandra-IRAC source separation is $\lesssim1.5$ arcsec. We also
estimate a fraction of spurious matches of randomized IRAC and X-ray
sources of $<1.5\%$ (i.e., $<$2 sources in our case) from the
cross-matching of IRAC and X-ray sources using a large offset in IRAC
coordinates ($3$ arcmin in either RA or Dec).

\subsection{Selection of IR power-law galaxies}
\label{subsectionSelectioniofIRplgalaxies}

Among all available IR-based colour selections of AGN, to construct
the IR power-law sample we used the IR colour-based selection of AGN
presented in \citet{Donley2012}, which has been optimized to be
  both effective and extremely reliable in identifying luminous AGN.
\citet{Donley2012} redefined the AGN selection criteria for deep IRAC
surveys using large samples of luminous AGN and high-redshift
star-forming galaxies in COSMOS.  Although this revised IRAC
  criterion is extremely sensitive to the reliability of the
  photometric estimates, it has been designed to minimize the
  contaminants, such as high-redshift star-forming galaxies, which are
  otherwise present in other high-redshift galaxy samples of IR-based
  AGN candidates \citep[e.g.,][]{Lacy2004,Stern2005}. \\

The revised AGN selection by \citet{Donley2012} is defined by the following wedge,

\begin{center}
\begin{gather*}
    x   =  \log{\left(\frac{f_{\nu,5.8\mu m}}{f_{\nu,3.6\mu m}}\right)} \quad , \quad 
    y   =  \log{\left(\frac{f_{\nu,8.0\mu m}}{f_{\nu,4.5\mu m}}\right)} \\
    x   \geq  0.08 \quad \text{and} \quad y \geq 0.15 \quad \text{and} \\ 
    y \geq  (1.21\times x)-0.27 \quad \text{and} \quad y \leq  (1.21\times x)+0.27 \quad \text{and}\\
    f_{\nu,3.6\mu m}  <   f_{\nu,4.5\mu m} < f_{\nu,5.8\mu m} < f_{\nu,8.0\mu m}
\end{gather*} 
\end{center}

Following these criteria, we can directly obtain an IR-based
classification: IR power-law are those lying within the revised
IRAC-selection wedge and their IRAC SEDs are monotonically-rising
within the IR bands. Thus, we label as IR non-power-law those galaxies,
which fall outside the \citet{Donley2012}'s wedge and all the sources
with non-monotonically-rising IRAC SEDs. The latter largely removes
any possible contamination due to low-redshift star-forming galaxies
(which would have little X-ray emission) in which their $1.6$ $\mu$m
stellar bump passes through the IRAC bandpass. Using these criteria,
the IR power-law sample contains $60$ X-ray-detected objects while the
IR non-power-law sample consists of $87$ X-ray-detected sources.\\

\begin{figure*}
    \centering
    \includegraphics[width=0.78\textwidth]{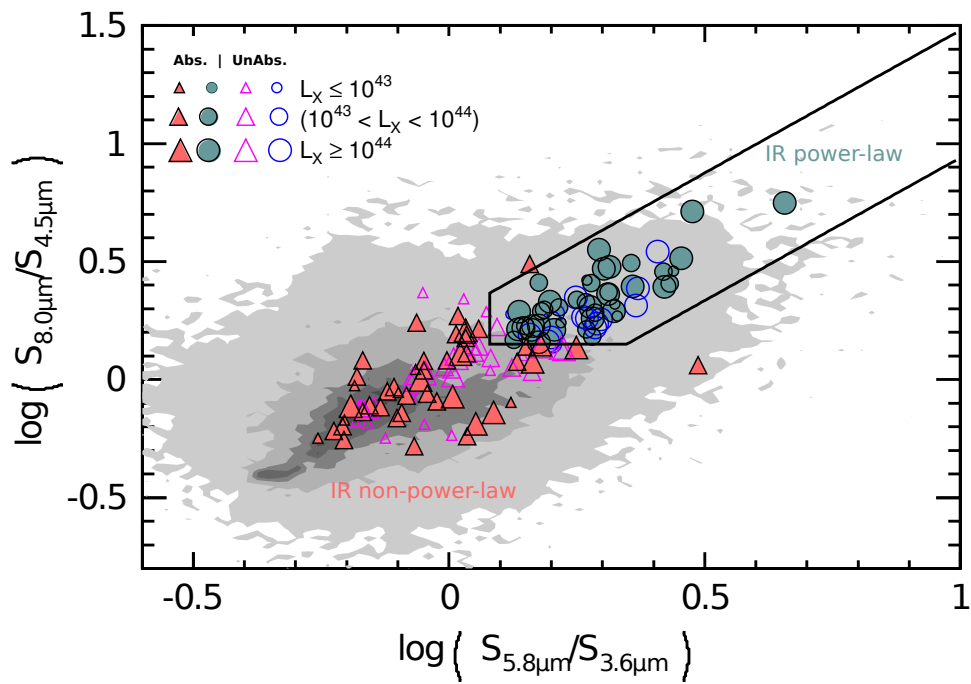}
    \caption{IRAC colour-colour diagram for the entire IRAC sample
      (plotted as a surface grey map) and for the cross-correlation
      sample: blue circles and red triangles represent IR power-law
      and IR non-power-law galaxies according to \citet{Donley2012},
      respectively. The filled and empty symbols denote absorbed and
      unabsorbed sources, respectively. The symbols change in size to
      denote different ranges of X-ray luminosity.  The solid line
      shows the revised IRAC-criteria wedge by \citet{Donley2012}.}
    \label{fig:IRACcolorcolor}
\end{figure*} 

Hereafter, we concentrate on the X-ray spectral properties of these
$147$  X-ray-detected galaxies, highlighting possible differences
between these two sub-samples (IR power-law and IR non-power-law). The
main aim of this work is to investigate their intrinsic absorption
distribution approaching the Compton-Thick limit and specially,
to infer whether obscured AGN are the major contributor to the IR
power-law-selected AGN survey or not, and specifically whether it is
effective at selecting type-2 AGN at some luminosity range. By
  design, we cannot address in this paper the question of the nature
  of the IR power-law X-ray-undetected sources.
Figure~\ref{fig:IRACcolorcolor} shows the IRAC colour distribution for
the entire IRAC sample.  The colour symbols are our $147$
X-ray-detected sources, with the blue circles showing the IR power-law
galaxies and the red triangles the IR non-power-law objects. Filled
and empty symbols denote absorbed ($N_{H}^{intr} > 10^{22}\, {\rm
  cm}^{-2}$) and unabsorbed ($N_{H}^{intr} \leq 10^{22}\, {\rm
  cm}^{-2}$) sources, respectively (see Section
\ref{intrinsicabsorption}). The symbols also change in size to denote
different ranges of the rest-frame 2-10 keV intrinsic luminosity.


\section{X-ray Spectral analysis}
\label{section3}

We have carried out an X-ray spectral analysis for the $147$ sources
whit background-subtracted EPIC counts in the 0.5-9 keV band above
$180$.  The \xmm~ data were grouped to have at least $20$ counts in
each bin in order to apply the modified $\chi^2$ minimization
technique.\\ The ability to obtain a reliable fit depends on the X-ray
spectral quality, or, in simpler terms, on the S/N of the X-ray
spectrum under analysis. The distribution of the net counts in the 2-9
keV band for all the sources in our sample peaks at $\sim$700 (see
Figure~\ref{fig:sample_properties}).  Despite the relatively high
mean value of the net counts in the hard energy band, there are many
cases in which the spectrum is dominated by the background. Therefore,
the strategy for the X-ray spectral analysis must be appropriate for
the high background regime.\\

A complete and detailed X-ray spectral analysis of the full sample of
X-ray sources in the CDFS {\it XMM-Newton} survey is beyond the scope
of this paper and will be presented in \citet{Comastri2013}.

\subsection{Spectral Models} 
\label{SpectralModels}

We carried out the spectral fitting of the $147$ sources using
\xspec{} v~12.7.0 \citep{Arnaud1996}.  We started with a joint fitting
of MOS and pn spectra with a power-law model (model {\tt A}, see
Table~\ref{tab:BestFittedModelSummary}) with fixed Galactic
absorption.  The Galaxy H column density values for each X-ray source
were obtained using the ftool
\emph{nh}\footnote{http://heasarc.nasa.gov/lheasoft/ftools}, which
interpolates over values from the HI map of \citet{Dickey1990}. The
MOS and pn spectral parameters were tied to the same value, while the
normalisations were left free to vary in order to account for flux
cross-calibration discrepancies between the EPIC MOS and pn cameras
\citep{Mateos2009}. An additional component was included in addition
to the power-law model to search for intrinsic rest-frame absorption:
a redshifted neutral absorption by cold matter with redshift fixed to
the source's spectroscopic or photometric value (model {\tt C}, see
Table~\ref{tab:BestFittedModelSummary}).  In a few cases ($5/147$),
the resulting photon index is $\lesssim1$ (even when accounting for
uncertainties), which is much lower than the typical values for
unabsorbed AGN.  A possible explanation would be that sources with
such an observed flat spectrum were Compton-Thick AGN in which all the
direct emission is suppressed and only reflected emission is observed
at energies below $10$ keV.  However, this explanation does not seem
to be valid because in all such cases.  We find a significant amount
of absorption but not in the Compton-Thick regime ($N_{H}^{intr}$'s
between $10^{22}$ and a few $10^{23}$ cm$^{-2}$). The iron K$\alpha$
emission line is not detected within sensitive limits ($>$3$\sigma$
level).  Our adopted explanation for the sources with an apparent very
flat X-ray spectral index is the combination of a moderate absorption
with poor statistics and/or high background (see, however,
\citet{Georgantopoulos2013} where a more thorough examination of these
sources yielded some Compton-Thick candidates).  In those cases, an
upper limit for the absorption is given at $90\%$ level by freezing
the spectral index at $1.9$, which is the average value for unabsorbed
AGN
\citep{Caccianiga2004,Mateos2005,Galbiati2005,Tozzi2006,Mateos2010}. \\

In other cases, the fit with a simple absorbed/unabsorbed power law is
not a good description of the X-ray spectrum.  Despite the fact that
the high background regime does not allow us to investigate more
complex spectral models as often present in AGN, we identify one
possible additional spectral component: a soft X-ray excess.  In this
paper we have adopted a simple parametrisation for this component and
we have only modelled it in the two simplest ways: one as a thermal
emission (model {\tt B} or {\tt D}, see
Table~\ref{tab:BestFittedModelSummary}) from a collisionally ionized
plasma, which is heated by shocks induced by AGN outflows
(e.g. \citealt{King2005}) or intense star formation
(e.g. \citealt{Schurch2002}); and the other one as a power-law-like
emission (model {\tt E}, see Table~\ref{tab:BestFittedModelSummary})
to account for the spectral complexity observed in some of our sources
and which may not be well fit by a simple thermal emission.\\

Strongly obscured AGN can show a reflection-dominated spectrum and/or
a high equivalent width iron line. These spectral features can be used
as additional criteria to signpost Compton-Thick AGN (as in e.g.
\citet{Georgantopoulos2013,Comastri2013}).  These kinds of
Compton-Thick AGN would not necessarily be identified unambiguously as
such in our analysis.\\

We performed the fits in the energy range 0.5-9 keV. We ignore the
data below $0.5$ keV to avoid uncertainties in the EPIC
calibration. The 1.4-1.6 keV energy range was ignored to avoid any
contribution from the Al-K fluorescence line in the internal \xmm~
EPIC background.  At high energies, the efficiency of \xmm~ decreases
rapidly and the energy bins $>9$ keV are dominated by noise and
background for most sources.\\

We measured the significance of the detection of additional components
in the X-ray spectra of our sources using the F-test statistic. The
F-test measures the significance of a decrease in $\chi^2$ when new
components are added to the model. We used significance thresholds of
$95\%$ to accept the detection of soft excess and/or intrinsic
absorption.  In three sources, the temperature of the soft X-ray
  excess component of the best-fit model is far too high (>10keV) to
  render this spectral component physically plausible. We consequently
  adopted as the best fit the one that yielded the least uncertain
  model parameters (provided that the values of these parameters are
  physically plausible). \\

\section{Spectral Results} 

\begin{figure}
    \centering
    \includegraphics[width=0.23\textwidth]{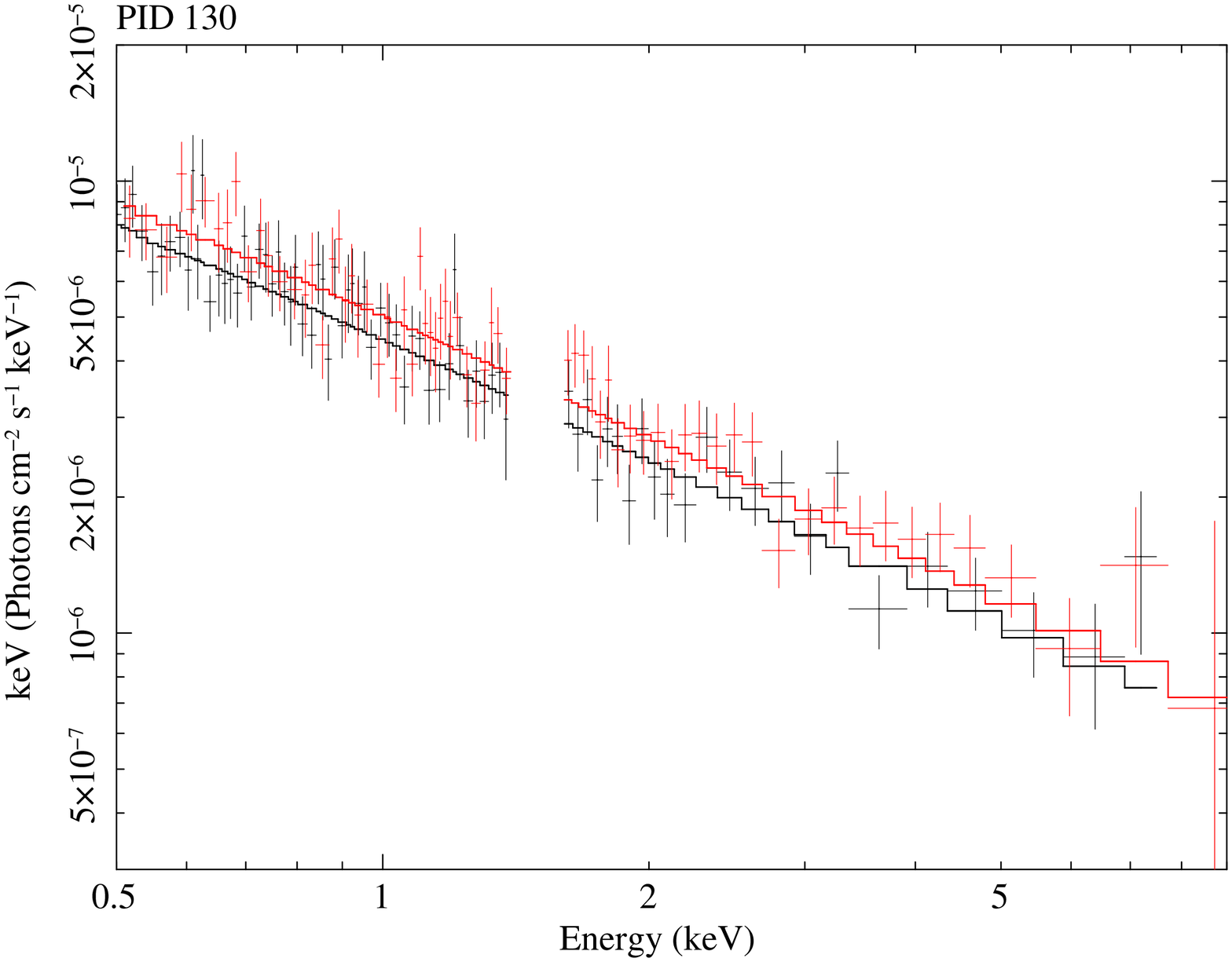}
    \includegraphics[width=0.23\textwidth]{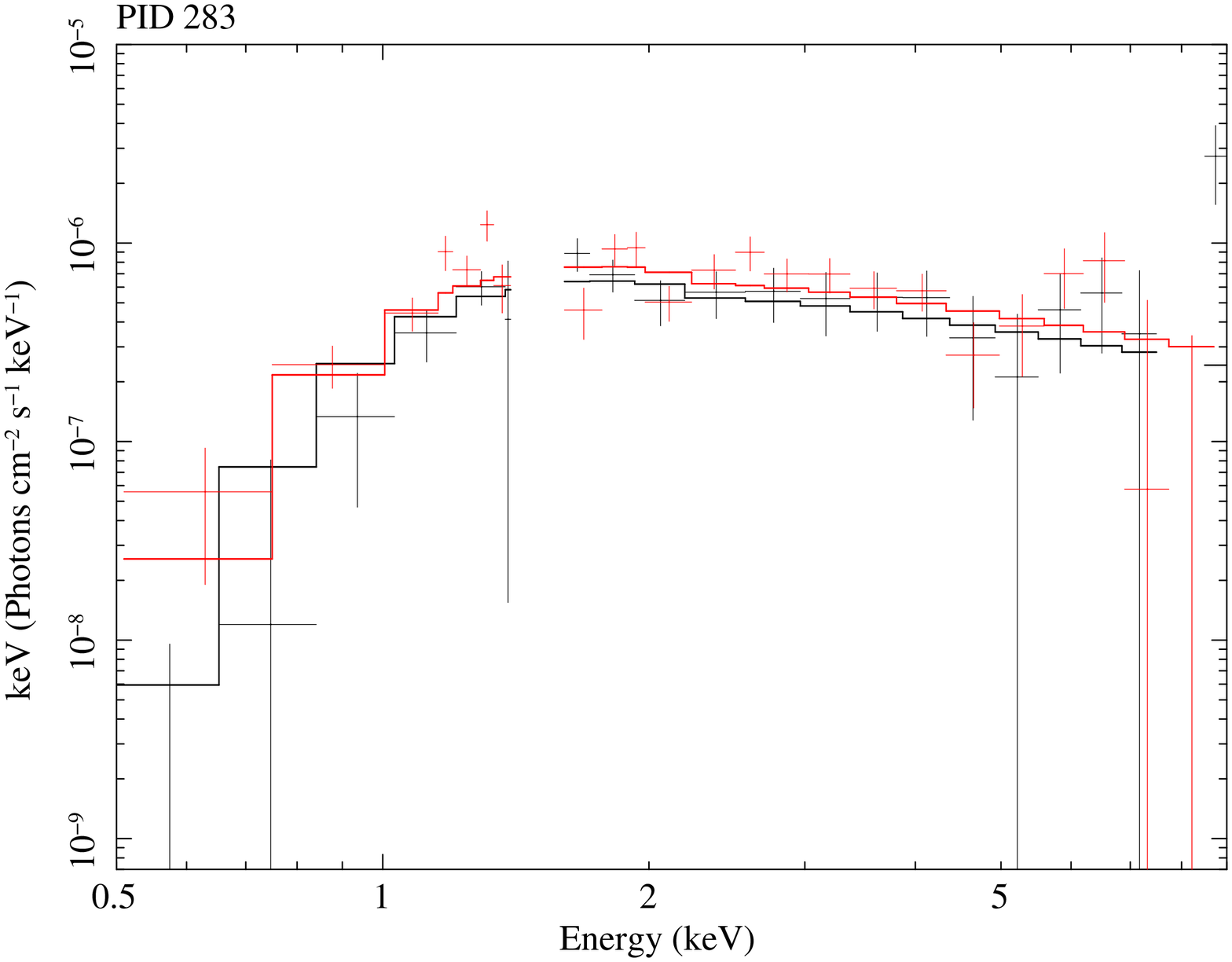}
    \includegraphics[width=0.23\textwidth]{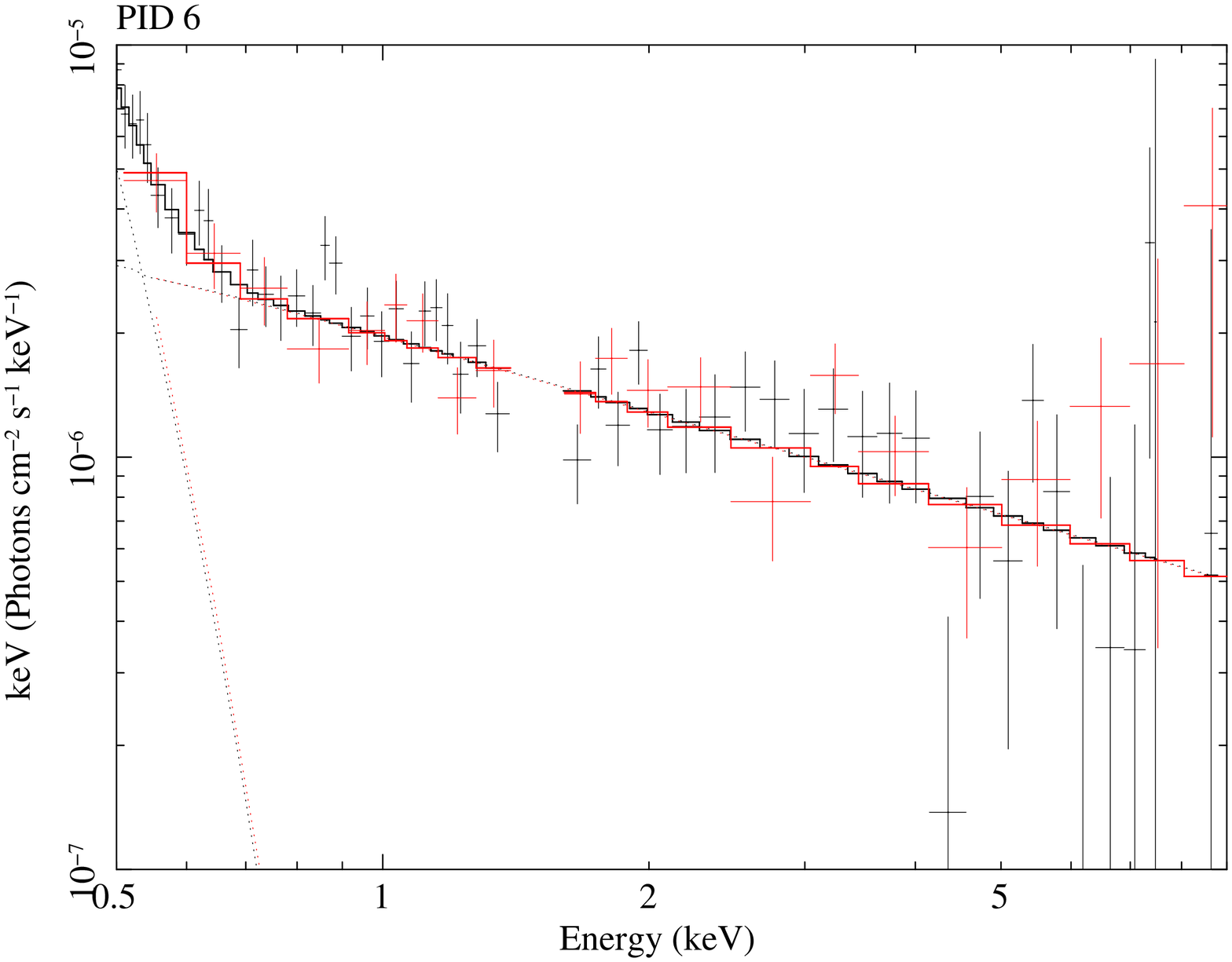}
    \includegraphics[width=0.23\textwidth]{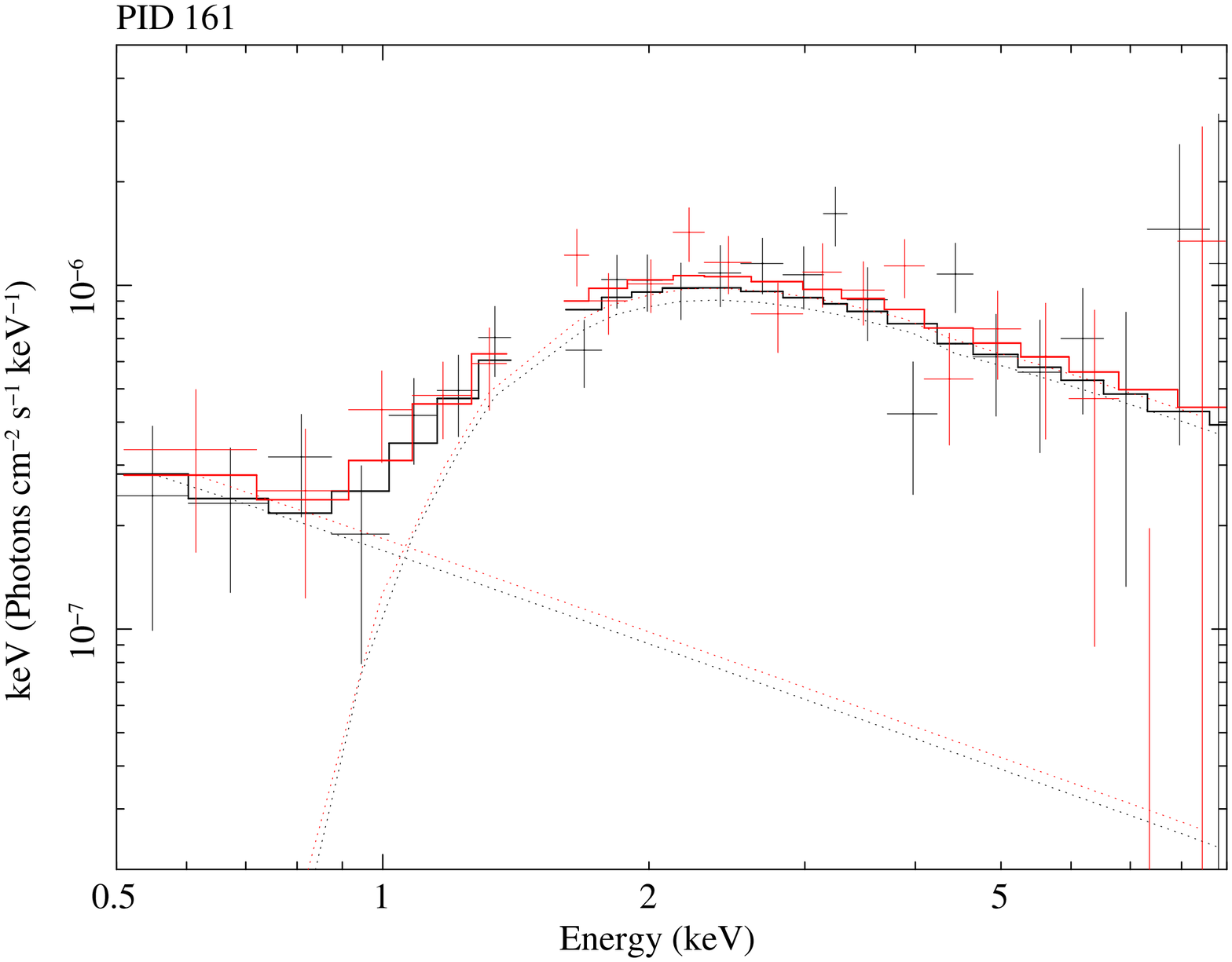}
    \caption{Example X-ray spectra of four objects whose best-fit
      model is a simple power law (model {\tt A}), absorbed power law
      (model {\tt C}), a simple power law with thermal soft excess
      (model {\tt B}) and an absorbed power law with a Thomson
      scattering component (model {\tt E}), respectively (ID210=130,
      ID210=283, ID210=6, ID210=161).}
    \label{fig:exampleFittedSpectrum}
\end{figure}

\begin{figure*}
    \centering
    \includegraphics[width=0.49\textwidth]{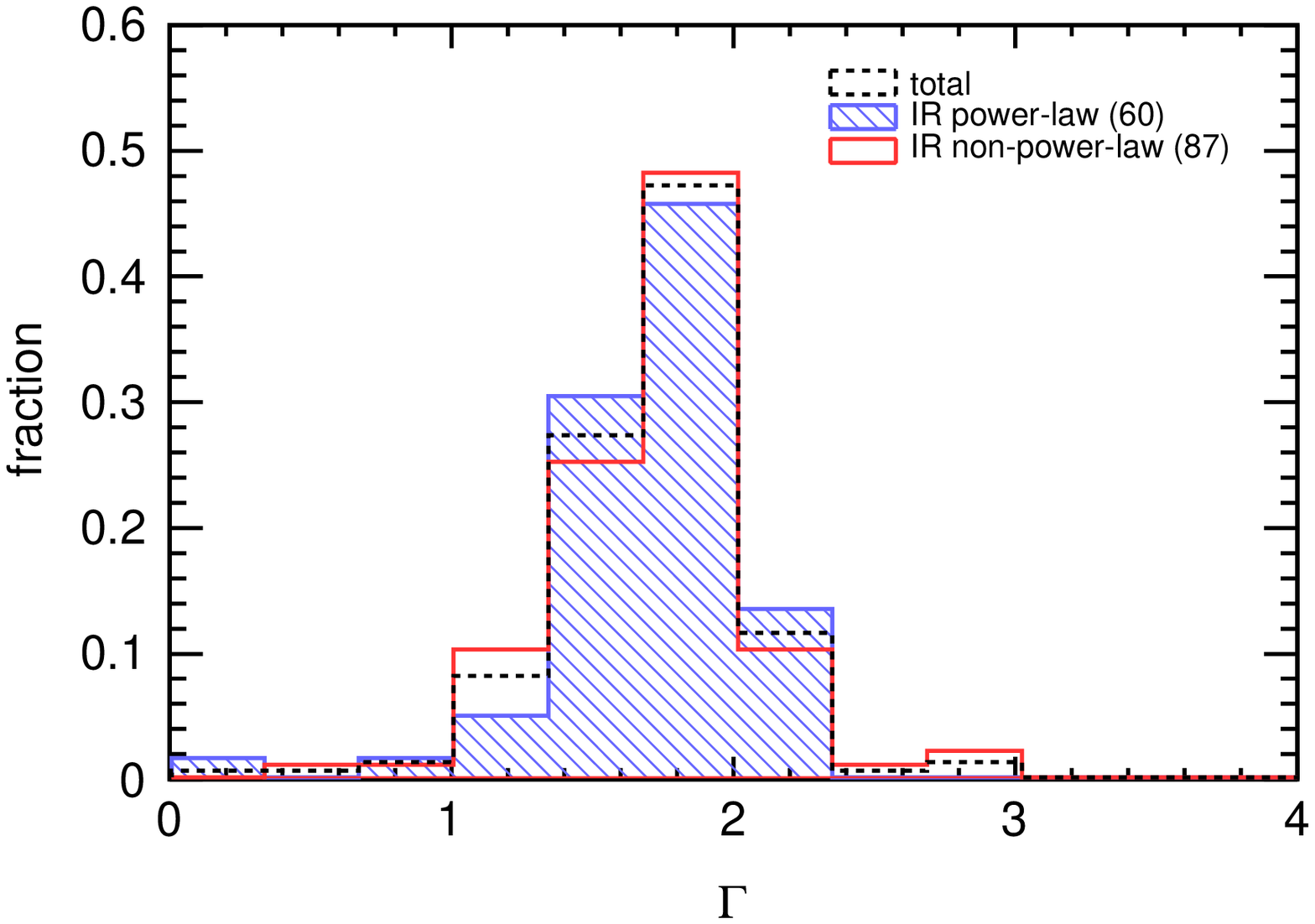}
    \includegraphics[width=0.49\textwidth]{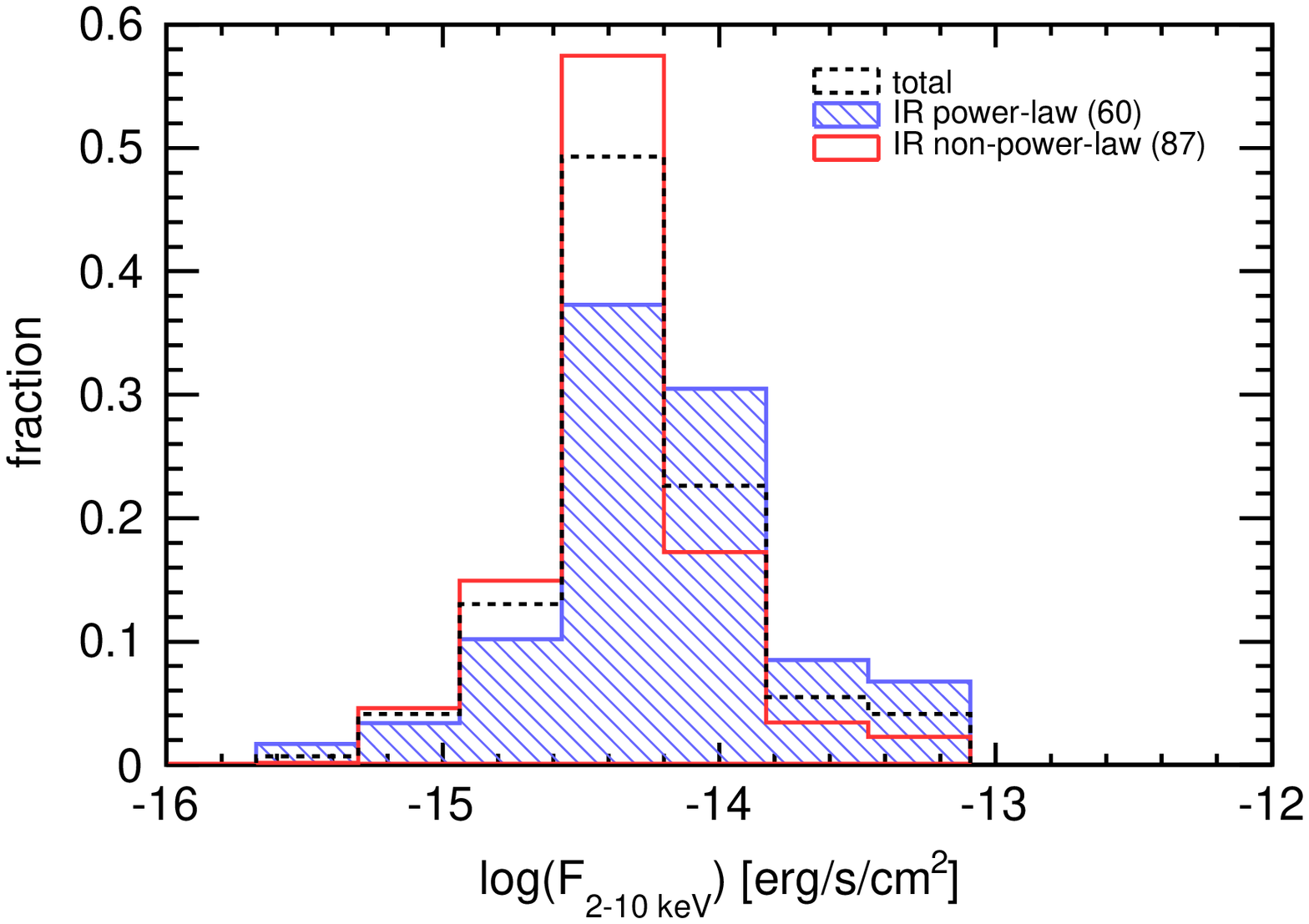}
    \includegraphics[width=0.49\textwidth]{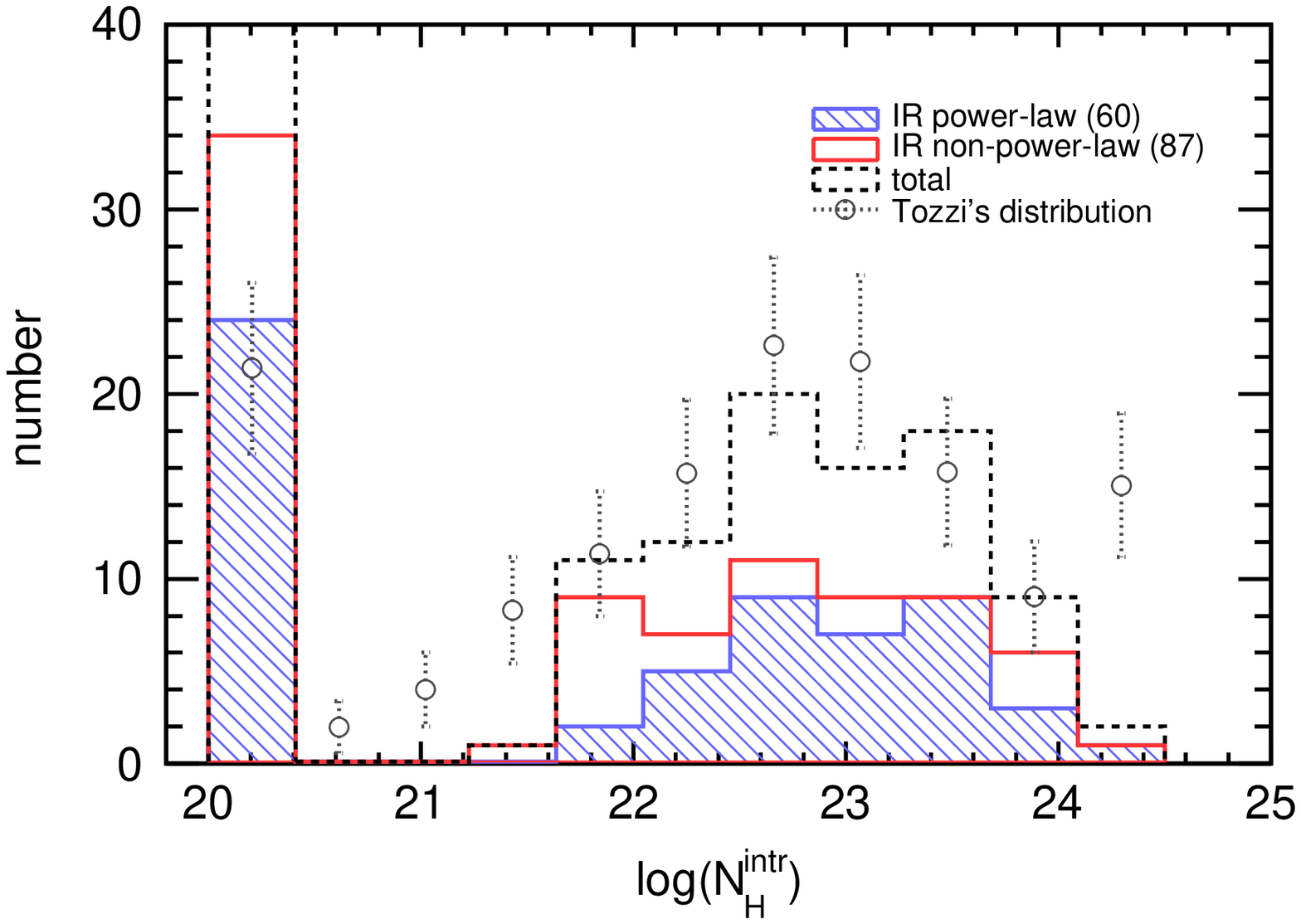}
    \includegraphics[width=0.49\textwidth]{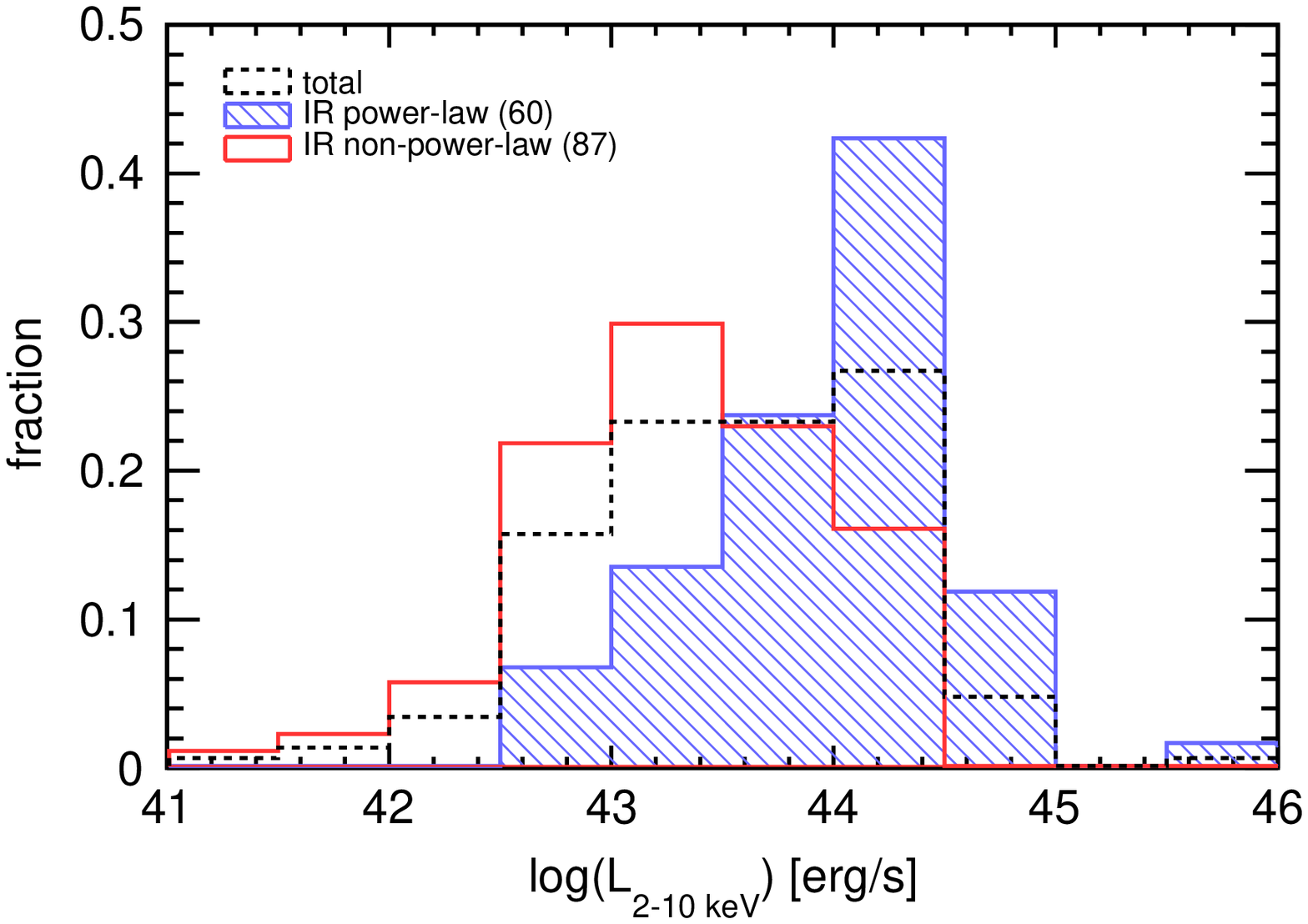}\\
    \caption{\emph{Top panels:} Normalised distribution of the X-ray
      power-law photon index and the observed 2-10 keV flux.
      \emph{Bottom panels:} Normalised distribution of the intrinsic
      absorption and the rest-frame 2-10 keV luminosity corrected for
      both Galactic and intrinsic absorption. The error points in the
      intrinsic absorption distribution show the $N_H^{intr}$
      distribution for the sample given by
      \citet{Tozzi2006}. Solid-line shaded and empty histograms show
      the distribution for the IR power-law and IR no power-law
      populations, respectively.  The dashed-line empty histogram
      shows the distribution for the entire sample.  All the values
      are given by the best fitted model (see
      Section~\ref{section3}). }
    \label{fig:fittedParameters}
\end{figure*}

Table~\ref{tab:Properties_of_the_sample} shows the best-fit model
parameters for each object and the overall properties of the whole
galaxy sample and the two subsamples are presented in
Table~\ref{tab:fittedParameters}.  A summary of the models required
during the spectral fit is summed up in
Table~\ref{tab:BestFittedModelSummary}.  We also show a examples of
fitted spectra for each component in
Figure~\ref{fig:exampleFittedSpectrum}.  Finally,
Figure~\ref{fig:fittedParameters} displays the distribution of some
model parameters of our sample, as computed from the best-fit model:
the rest-frame X-ray power-law photon index, the intrinsic absorption
column density, the observed 2-10 keV fluxes, and the 2-10 keV
rest-frame luminosity corrected for both intrinsic and Galactic
absorption. Whilst the solid-line hatched and empty histograms show
the distribution for IR power-law and IR non-power-law galaxies,
respectively, the dashed-line empty histogram shows the distribution
for the whole sample. As we can see in the Figure, It is clear that IR
power-law galaxies are distinct among the whole sample in several
respects, as we shall see in the following subsections.\\

\begin{table*}
	\centering
	\rotatebox{90}{
	\begin{minipage}{\textheight}
	\caption{Details of the X-ray sources with IRAC counterpart
          found in this analysis presente in three panels. Column (1)
          is the identification number of X-ray sources listed in
          Ranalli et al. (in prep.); column (2) is the redshift of the
          source (rounded to two (photometric) or three
          (spectroscopic) decimal); column (3) gives the observed
          $2-10$ keV flux of the source in units of $10^{-15}$
          erg/s/cm$^{2}$; column (4) gives the unabsorbed rest-frame
          $2-10$ keV luminosity of the source using the best fit model
          in units of $10^{42}$ erg/s; column (5) is the X-ray photon
          index, $\Gamma$, where $\dagger$ indicates that the index
          was frozen in the fit; column (6) is the intrinsic column 
          density $N_H$ measured from the best fit model in column (8)
          in units $10^{22}$ cm$^{-2}$; column (7) gives the electron
          temperature for the soft Comptonisation component in units
          of keV; column (8) is the best-fitting model for the
          spectrum as described in Section \ref{SpectralModels}, see
          also Table \ref{tab:BestFittedModelSummary}; column (9) is a
          tick to denote whether the sources is classified as IR
          power-law or IR no-power-law galaxy; (10) 1$\sigma$ confidence 
          interval on the intrinsic column density $N_H^{intr}$ measured using 
          models {\tt C}, {\tt D}, {\tt E} in units of $10^{22}$ cm$^{-2}$ 
          (see Section 5); (11) classification of the source as absorbed, 
          unabsorbed and unclassified as $a$, $u$ and $-$, 
          respectively (see Section 5). }
	\label{tab:Properties_of_the_sample}
	\resizebox{\textwidth}{7cm}{ %
    \footnotesize{
	\begin{tabular}{ | ccccccccccc | ccccccccccc | ccccccccccc |}\hline
 PID &  z & $F_{X}$  &  $L_{X}$  &  $\Gamma$  &  $N_{H}^{intr}$  &  $KT$  &  model & IRpl &   $N_H^{intr} CL(1\sigma)$ & class & 
 PID &  z & $F_{X}$  &  $L_{X}$  &  $\Gamma$  &  $N_{H}^{intr}$  &  $KT$  &  model & IRpl &   $N_H^{intr} CL(1\sigma)$ & class & 
 PID &  z & $F_{X}$  &  $L_{X}$  &  $\Gamma$  &  $N_{H}^{intr}$  &  $KT$  &  model & IRpl &   $N_H^{intr} CL(1\sigma)$ & class \\
 (1) & (2) & (3) & (4) & (5) & (6) & (7) & (8) & (9) & (10) & (11) & 
 (1) & (2) & (3) & (4) & (5) & (6) & (7) & (8) & (9) & (10) & (11) & 
 (1) & (2) & (3) & (4) & (5) & (6) & (7) & (8) & (9) & (10) & (11) \\\hline\hline
 2 &  $1.622$  &  $9.89$  &  $195.68$  & $2.1\pm^{0.1}_{0.1}$ &  -  &  -  & {\tt A} & \ding{51} &  $0.02-0.17$  &  $u$ &                                                              116 &  $3.53$  &  $4.91$  &  $475.68$  & $1.8\pm^{0.3}_{0.2}$ &  $43\pm^{16}_{14}$ &  -  & {\tt E} & \ding{51} &  $22-38$  &  $a$ &                                234 &  $0.952$  &  $3.11$  &  $12.45$  & $1.8\pm^{0.2}_{0.2}$ &  $1.04\pm^{0.53}_{0.45}$ &  -  & {\tt C} & \ding{55} &  $0.91-1.2$  &  $-$ \\ 
 3 &  $1.59$  &  $4.20$  &  $64.01$  & $1.9\pm^{0.4}_{0.4}$ &  $8.37\pm^{3.83}_{2.84}$ &  -  & {\tt C} & \ding{55} &  $7-11$  &  $a$ &                                               118 &  $1.609$  &  $8.60$  &  $172.30$  & $2.2\pm^{0.1}_{0.1}$ &  $2.40\pm^{0.27}_{0.25}$ &  -  & {\tt C} & \ding{51} &  $2.1-2.7$  &  $a$ &                        237 &  $0.620$  &  $4.10$  &  $6.09$  & $1.8\pm^{0.1}_{0.1}$ &  -  &  -  & {\tt A} & \ding{55} &  $\geq0.24$  &  $-$ \\
 6 &  $0.526$  &  $9.15$  &  $8.48$  & $1.6\pm^{0.1}_{0.1}$ &  -  &  $0.067\pm^{0.024}_{0.021}$ & {\tt B} & \ding{51} &  $\leq0.04$  &  $u$ &                                         120 &  $3.591$  &  $3.31$  &  $217.70$  & $1.6\pm^{0.1}_{0.1}$ &  -  &  -  & {\tt A} & \ding{55} &  $1.3-2.4$  &  $a$ &                                             244 &  $1.260$  &  $4.45$  &  $37.87$  & $1.9\pm^{0.1}_{0.1}$ &  -  &  -  & {\tt A} & \ding{55} &  $0.054-0.18$  &  $u$ \\
13 &  $1.05$  &  $4.19$  &  $16.42$  & $1.4\pm^{0.4}_{0.3}$ &  $0.86\pm^{0.96}_{0.75}$ &  -  & {\tt C} & \ding{55} &  $0.63-1.10$  &  $-$ &                                          123 &  $1.43$  &  $2.93$  &  $29.43$  & $1.7\pm^{0.6}_{0.5}$ &  $2.34\pm^{2.56}_{1.77}$ &  -  & {\tt C} & \ding{55} &  $1.8-3.0$  &  $a$ &                         245 &  $1.864$  &  $1.01$  &  $6.88$  & $\leq1.52$ &  $27\pm^{43}_{17}$ &  -  & {\tt C} & \ding{55} &  $41-150$  &  $a$ \\
19 &  $0.662$  &  $6.00$  &  $10.16$  & $1.7\pm^{0.3}_{0.3}$ &  $1.57\pm^{0.71}_{0.56}$ &  -  & {\tt C} & \ding{55} &  $1-3$  &  $a$ &                                                124 &  $0.35$  &  $1.39$  &  $0.53$  & $1.8\pm^{0.2}_{0.1}$ &  -  &  -  & {\tt A} & \ding{55} &  $\geq0.24$  &  $-$ &                                              248 &  $0.922$  &  $0.77$  &  $2.39$  & $1.5\pm^{0.3}_{0.3}$ &  -  &  -  & {\tt A} & \ding{55} &  $0.18-0.53$  &  $-$ \\
21 &  $1.156$  &  $3.68$  &  $22.15$  & $1.7\pm^{0.6}_{0.5}$ &  $3.04\pm^{2.86}_{1.81}$ &  -  & {\tt C} & \ding{55} &  $1-6$  &  $a$ &                                                130 &  $1.628$  &  $13.44$ &  $212.81$  & $1.9\pm^{0.0}_{0.0}$ &  -  &  -  & {\tt A} & \ding{51} &  $\leq0.04$  &  $u$ &                                            249 &  $0.738$  &  $7.35$  &  $18.99$  & $2.1\pm^{0.1}_{0.1}$ &  -  &  $0.075\pm^{0.042}_{0.025}$ & {\tt B} & \ding{55} &  $\geq0.24$  &  $-$ \\
24 &  $1.368$  &  $2.67$  &  $34.46$  & $1.9\pm^{1.7}_{1.1}$ &  $29\pm^{27}_{15}$ &  -  & {\tt C} & \ding{55} &  $20-50$  &  $a$ &                                                    133 &  $0.671$  &  $2.06$  &  $3.62$  & $1.8\pm^{0.1}_{0.1}$ &  -  &  -  & {\tt A} & \ding{55} &  $\leq0.08$  &  $u$ &                                              262 &  $1.621$  &  $8.61$  &  $134.48$  & $1.9\pm^{0.0}_{0.0}$ &  -  &  -  & {\tt A} & \ding{51} &  $\leq0.04$  &  $u$ \\
26 &  $3.198$  &  $2.73$  &  $239.94$  & $1.9\pm^{1.2}_{0.8}$ &  $46\pm^{60}_{29}$ &  -  & {\tt C} & \ding{51} &  $28-70$  &  $a$ &                                                   134 &  $1.609$  &  $16.81$ &  $224.82$  & $1.8\pm^{0.0}_{0.0}$ &  -  &  -  & {\tt A} & \ding{51} &  $\leq0.05$  &  $u$ &                                            265 &  $0.665$  &  $3.89$  &  $5.80$  & $1.5\pm^{0.1}_{0.1}$ &  -  & $0.070\pm^{0.080}_{0.030}$ & {\tt B} & \ding{55} &  $0.09-0.18$  &  $u$ \\
30 &  $1.94$  &  $9.86$  &  $204.64$  & $1.5\pm^{0.8}_{0.6}$ &  $77\pm^{60}_{36}$ &  -  & {\tt C} & \ding{55} &  $400-1400$  &  $a$ &                                                137 &  $0.751$  &  $1.82$  &  $6.00$  & $2.3\pm^{1.1}_{0.8}$ &  $4.50\pm^{3}_{2}$ & $0.067\pm^{0.05}_{0.04}$ & {\tt D} & \ding{55} &  $2.3-5.2$  &  $a$ &           269 &  $1.76$  &  $4.74$  &  $66.75$  & $1.6\pm^{0.3}_{0.3}$ &  $2.32\pm^{1.71}_{1.36}$ &  -  & {\tt C} & \ding{51} &  $1.9-2.8$  &  $a$ \\
31 &  $2.583$  &  $7.67$  &  $176.62$  & $1.3\pm^{0.1}_{0.1}$ &  -  &  -  & {\tt A} & \ding{51} &  $3-7$  &  $a$ &                                                                    138 &  $1.220$  &  $3.44$  &  $23.26$  & $1.7\pm^{0.1}_{0.1}$ &  -  &  -  & {\tt A} & \ding{55} &  $\leq0.08$  &  $u$ &                                             273 &  $0.733$  &  $4.97$  &  $11.38$  & $1.9\pm^{0.1}_{0.1}$ &  -  &  -  & {\tt A} & \ding{55} &  $\leq0.008$  &  $u$ \\
33 &  $1.843$  &  $35.02$  &  $637.08$  & $1.7\pm^{0.0}_{0.1}$ &  -  &  $0.224\pm^{0.072}_{0.068}$ & {\tt B} & \ding{51} &  $2-4$  &  $a$ &                                           142 &  $1.38$  &  $3.19$  &  $19.82$  & $1.2\pm^{0.5}_{0.4}$ &  $10\pm^{8}_{6}$ &  -  & {\tt C} & \ding{55} &  $13-26$  &  $a$ &                                   277 &  $1.323$  &  $6.71$  &  $69.56$  & $2.0\pm^{0.1}_{0.1}$ &  -  &  -  & {\tt A} & \ding{55} &  $\leq0.06$  &  $u$ \\
34 &  $1.63$  &  $1.13$  &  $9.08$  & $1.9\dagger$ &  $82\pm^{36}_{30}$  &  -  & {\tt C} & \ding{51} &  $500-1200$  &  $a$ &                                                         144 &  $3.700$  &  $0.93$  &  $5592.82$  & $\leq4$ &  $249\pm^{89}_{132}$ & $0.469\pm^{0.400}_{0.269}$ & {\tt D} & \ding{51} &  $170-350$  &  $a$ &                 281 &  $1.26$  &  $8.42$  &  $69.26$  & $1.8\pm^{0.1}_{0.1}$ &  $0.86\pm^{0.22}_{0.21}$ &  -  & {\tt C} & \ding{55} &  $0.80-0.92$  &  $-$ \\
37 &  $0.624$  &  $3.75$  &  $10.21$  & $1.9\dagger$ &  $20\pm^{11}_{8}$  &  -  & {\tt C} & \ding{55} &  $10-30$  &  $a$ &                                                            146 &  $0.122$  &  $8.24$  &  $0.33$  & $1.6\pm^{0.2}_{0.2}$ &  $1.79\pm^{0.33}_{0.29}$ &  -  & {\tt C} & \ding{55} &  $1.5-2.1$  &  $a$ &                          283 &  $2.291$  &  $4.46$  &  $147.98$  & $1.8\pm^{0.3}_{0.3}$ &  $18\pm^{6}_{5}$ &  -  & {\tt C} & \ding{51} &  $13-24$  &  $a$ \\
38 &  $1.598$  &  $9.37$  &  $129.04$  & $1.8\pm^{0.1}_{0.1}$ &  -  &  -  & {\tt A} & \ding{51} &  $\geq0.24$  &  $-$ &                                                               147 &  $1.536$  &  $5.85$  &  $58.53$  & $1.3\pm^{0.7}_{0.6}$ &  $59\pm^{42}_{32}$ &  -  & {\tt C} & \ding{55} &  $27-100$  &  $a$ &                                284 &  $0.566$  &  $4.51$  &  $4.80$  & $1.5\pm^{0.2}_{0.2}$ &  -  &  $0.218\pm^{0.038}_{0.048}$ & {\tt B} & \ding{55} &  $---$  &  $-$ \\
40 &  $0.512$  &  $4.58$  &  $4.43$  & $1.9\pm^{0.1}_{0.1}$ &  -  &  -  & {\tt A} & \ding{55} &  $0.02-0.11$  &  $u$ &                                                                151 &  $1.51$  &  $7.31$  &  $74.73$  & $1.6\pm^{0.1}_{0.1}$ &  $4.22\pm^{0.92}_{0.82}$ &  -  & {\tt C} & \ding{51} &  $3.4-5.1$  &  $a$ &                         285 &  $2.072$  &  $3.58$  &  $101.20$  & $1.9\pm^{0.1}_{0.1}$ &  -  &  -  & {\tt A} & \ding{55} &  $\geq0.24$  &  $-$ \\
42 &  $0.619$  &  $3.01$  &  $4.71$  & $1.9\pm^{0.2}_{0.2}$ &  $0.21\pm^{0.17}_{0.16}$ &  -  & {\tt C} & \ding{55} &  $0.05-0.38$  &  $u$ &                                           155 &  $0.732$  &  $3.23$  &  $6.39$  & $1.3\pm^{1.0}_{1.0}$ &  $18\pm^{22}_{15}$ &  -  & {\tt C} & \ding{55} &  $17-44$  &  $a$ &                                  289 &  $0.607$  &  $11.70$  &  $12.87$  & $1.2\pm^{0.2}_{0.2}$ &  $4.18\pm^{2.56}_{1.74}$ &  -  & {\tt E} & \ding{55} &  $0.97-1.30$  &  $-$ \\
43 &  $1.167$  &  $3.89$  &  $24.80$  & $1.8\pm^{0.1}_{0.1}$ &  -  &  -  & {\tt A} & \ding{55} &  $\leq0.05$  &  $u$ &                                                                158 &  $2.394$  &  $4.54$  &  $141.01$  & $1.6\pm^{0.3}_{0.3}$ &  $22\pm^{8}_{7}$ &  -  & {\tt C} & \ding{51} &  $16-30$  &  $a$ &                                  291 &  $1.215$  &  $3.30$  &  $26.54$  & $1.6\pm^{1.3}_{1.0}$ &  $39\pm^{50}_{26}$ &  -  & {\tt C} & \ding{51} &  $28-77$  &  $a$ \\
44 &  $1.51$  &  $3.45$  &  $22.13$  & $1.1\pm^{0.3}_{0.3}$ &  -  &  -  & {\tt A} & \ding{55} &  $\geq0.03$  &  $-$ &                                                                161 &  $0.686$  &  $7.71$  &  $17.00$  & $1.9\pm^{0.3}_{0.3}$ &  $7.10\pm^{2.44}_{2.07}$ &  -  & {\tt E} & \ding{55} &  $5.0-9.5$  &  $a$ &                         297 &  $1.44$  &  $7.54$  &  $118.57$  & $2.0\pm^{0.6}_{0.5}$ &  $30\pm^{16}_{11}$ &  -  & {\tt C} & \ding{51} &  $20-46$  &  $a$ \\
45 &  $0.717$  &  $4.23$  &  $7.67$  & $1.6\pm^{0.1}_{0.1}$ &  -  &  -  & {\tt A} & \ding{55} &  $0.10-0.24$  &  $u$ &                                                                165 &  $1.016$  &  $4.22$  &  $17.69$  & $1.5\pm^{0.3}_{0.3}$ &  $6.49\pm^{2.37}_{1.91}$ & $0.069\pm^{0.069}_{0.052}$ & {\tt D} & \ding{55} &  $7.0-14$  &  $a$ &   301 &  $0.665$  &  $2.25$  &  $3.72$  & $1.7\pm^{0.1}_{0.1}$ &  -  &  -  & {\tt A} & \ding{55} &  $0.07-0.20$  &  $u$ \\
48 &  $0.298$  &  $42.64$  &  $11.94$  & $1.5\pm^{0.1}_{0.1}$ &  $4.36\pm^{0.81}_{0.62}$ &  $0.242\pm^{0.081}_{0.049}$ & {\tt D} & \ding{55} &  $4.6-5.7$  &  $a$ &                   166 &  $1.36$  &  $1.90$  &  $36.06$  & $2.3\pm^{1.7}_{1.0}$ &  $40\pm^{66}_{25}$ &  -  & {\tt C} & \ding{55} &  $19-51$  &  $a$ &                                 305 &  $0.467$  &  $28.99$  &  $20.75$  & $1.6\pm^{0.1}_{0.1}$ &  $0.46\pm^{0.34}_{0.22}$ &  $0.090\pm^{0.052}_{0.056}$ & {\tt D} & \ding{51} &  $1.2-2.2$  &  $a$ \\
49 &  $2.298$  &  $4.55$  &  $103.95$  & $1.5\pm^{0.3}_{0.2}$ &  $3.43\pm^{2.25}_{1.86}$ &  -  & {\tt C} & \ding{51} &  $1.6-5.7$  &  $a$ &                                           173 &  $1.920$  &  $4.78$  &  $100.08$  & $1.8\pm^{0.1}_{0.1}$ &  -  &  -  & {\tt A} & \ding{55} &  $\leq0.24$  &  $u$ &                                            307 &  $0.857$  &  $5.53$  &  $16.76$  & $1.7\pm^{0.3}_{0.3}$ &  $1.06\pm^{0.60}_{0.48}$ &  -  & {\tt C} & \ding{55} &  $0.9-1.2$  &  $-$ \\
50 &  $1.218$  &  $3.65$  &  $26.88$  & $1.8\pm^{0.3}_{0.3}$ &  $2.16\pm^{1.23}_{0.95}$ &  -  & {\tt C} & \ding{55} &  $1.2-3.4$  &  $a$ &                                            174 &  $1.581$  &  $13.85$  &  $180.35$  & $1.8\pm^{0.1}_{0.1}$ &  -  &  -  & {\tt A} & \ding{55} &  $\geq0.24$  &  $-$ &                                           308 &  $0.736$  &  $13.52$  &  $30.09$  & $1.8\pm^{0.0}_{0.0}$ &  -  &  -  & {\tt A} & \ding{55} &  $0.01-0.06$  &  $u$ \\
57 &  $2.571$  &  $2.71$  &  $184.29$  & $2.2\pm^{0.1}_{0.1}$ &  -  &  -  & {\tt A} & \ding{51} &  $\leq0.08$  &  $u$ &                                                               175 &  $1.52$  &  $5.04$  &  $57.70$  & $1.6\pm^{0.4}_{0.3}$ &  $24\pm^{10}_{7}$ &  -  & {\tt C} & \ding{51} &  $17-34$  &  $a$ &                                  310 &  $0.668$  &  $10.21$  &  $20.66$  & $1.9\pm^{0.2}_{0.2}$ &  $6.93\pm^{1.50}_{1.38}$ &  -  & {\tt E} & \ding{51} &  $5.6-8.4$  &  $a$ \\
60 &  $1.049$  &  $3.78$  &  $17.66$  & $1.6\pm^{0.4}_{0.3}$ &  $4.24\pm^{1.73}_{1.42}$ &  -  & {\tt C} & \ding{55} &  $2.8-6.0$  &  $a$ &                                            178 &  $1.380$  &  $1.24$  &  $19.47$  & $2.3\pm^{0.4}_{0.3}$ &  $1.11\pm^{0.73}_{0.61}$ &  -  & {\tt C} & \ding{51} &  $0.92-1.3$  &  $-$ &                        311 &  $2.091$  &  $9.34$  &  $211.98$  & $1.7\pm^{0.2}_{0.1}$ &  $6.36\pm^{1.46}_{1.31}$ &  -  & {\tt C} & \ding{51} &  $5.1-7.8$  &  $a$ \\
62 &  $2.561$  &  $6.51$  &  $369.28$  & $2.0\pm^{0.1}_{0.1}$ &  -  &  $0.278\pm^{0.119}_{0.130}$  & {\tt B} & \ding{51} &  $14-48$  &  $a$ &                                         180 &  $3.064$  &  $5.43$  &  $376.04$  & $1.8\pm^{0.4}_{0.3}$ &  $87\pm^{28}_{23}$ &  -  & {\tt C} & \ding{51} &  $65-110$  &  $a$ &                               312 &  $1.406$  &  $7.44$  &  $65.37$  & $1.6\pm^{0.3}_{0.2}$ &  $7.38\pm^{3.33}_{2.22}$ &  -  & {\tt E} & \ding{51} &  $2.0-4.4$  &  $a$ \\
64 &  $3.34$  &  $6.75$  &  $327.35$  & $1.4\pm^{0.6}_{0.4}$ &  $70\pm^{44}_{31}$ &  -  & {\tt C} & \ding{51} &  $39-110$  &  $a$ &                                                  182 &  $1.034$  &  $4.96$  &  $22.11$  & $1.7\pm^{0.1}_{0.1}$ &  -  &  -  & {\tt A} & \ding{55} &  $\leq0.007$  &  $u$ &                                            315 &  $0.534$  &  $2.83$  &  $2.91$  & $1.8\pm^{0.1}_{0.1}$ &  -  &  -  & {\tt A} & \ding{55} &  $\geq0.24$  &  $-$ \\
66 &  $1.185$  &  $6.25$  &  $112.48$  & $1.9\pm^{1.7}_{1.5}$ &  $143\pm^{139}_{101}$ &  -  & {\tt C} & \ding{55} &  $43-280$  &  $a$ &                                               185 &  $0.735$  &  $1.65$  &  $4.03$  & $2.0\pm^{0.4}_{0.3}$ &  $0.59\pm^{0.43}_{0.35}$ &  -  & {\tt C} & \ding{55} &  $0.49-2.71$  &  $-$ &                        318 &  $0.663$  &  $6.12$  &  $11.30$  & $1.9\pm^{0.2}_{0.2}$ &  -  &  $0.186\pm^{0.047}_{0.066}$ & {\tt B} & \ding{55} &  $\geq0.24$  &  $-$ \\
68 &  $2.005$  &  $6.99$  &  $173.27$  & $1.9\pm^{0.1}_{0.1}$ &  -  &  -  & {\tt A} & \ding{55} &  $\geq0.3$  &  $-$ &                                                                190 &  $3.10$  &  $3.75$  &  $204.51$  & $1.7\pm^{0.2}_{0.1}$ &  $2.48\pm^{1.72}_{1.55}$ &  -  & {\tt C} & \ding{55} &  $2.0-3.0$  &  $a$ &                        319 &  $0.742$  &  $80.80$  &  $189.17$  & $1.9\pm^{0.0}_{0.0}$ &  -  &  $0.214\pm^{0.018}_{0.020}$ & {\tt B} & \ding{51} &  $\geq0.24$  &  $-$ \\
71 &  $0.494$  &  $3.99$  &  $4.53$  & $2.3\pm^{0.6}_{0.4}$ &  $2.30\pm^{3.47}_{0.92}$ &  $0.135\pm^{0.712}_{0.098}$ & {\tt D} & \ding{51} &  $2.2-4.1$  &  $a$ &                     191 &  $0.821$  &  $6.07$  &  $14.90$  & $1.6\pm^{0.1}_{0.1}$ &  -  &  -  & {\tt A} & \ding{55} &  $0.04-0.19$  &  $u$ &                                            320 &  $0.735$  &  $3.69$  &  $4.64$  & $\leq1.87$ &  $9.86\pm^{18.25}_{6.51}$ &  -  & {\tt C} & \ding{55} &  $17-80$  &  $a$ \\
72 &  $0.839$  &  $3.31$  &  $11.09$  & $2.0\pm^{0.2}_{0.2}$ &  $1.47\pm^{0.52}_{0.43}$ &  -  & {\tt C} & \ding{55} &  $1.0-2.0$  &  $a$ &                                            194 &  $2.838$  &  $3.64$  &  $219.13$  & $1.9\pm^{0.2}_{0.2}$ &  $28\pm^{7}_{6}$ &  -  & {\tt C} & \ding{51} &  $22-35$  &  $a$ &                                  321 &  $0.733$  &  $5.77$  &  $8.18$  & $1.0\pm^{0.1}_{0.1}$ &  -  &  -  & {\tt A} & \ding{55} &  $\geq0.24$  &  $-$ \\
78 &  $0.737$  &  $17.06$  &  $44.77$  & $1.7\pm^{0.3}_{0.3}$ &  $16\pm^{4}_{3}$ &  -  & {\tt C} & \ding{55} &  $12-20$  &  $a$ &                                                     197 &  $0.667$  &  $1.62$  &  $2.48$  & $1.5\pm^{0.4}_{0.3}$ &  $0.91\pm^{0.85}_{0.60}$ &  -  & {\tt C} & \ding{55} &  $0.7-1.1$  &  $-$ &                          323 &  $0.181$  &  $2.95$  &  $0.25$  & $1.4\pm^{0.2}_{0.2}$ &  -  &  -  & {\tt A} & \ding{55} &  $\leq0.03$  &  $u$ \\
81 &  $1.887$  &  $17.70$  &  $311.20$  & $1.6\pm^{0.1}_{0.1}$ &  -  &  $0.411\pm^{0.059}_{0.077}$ & {\tt B} & \ding{51} &  $\geq0.24$  &  $-$ &                                      198 &  $0.860$  &  $5.82$  &  $32.79$  & $2.2\pm^{0.8}_{0.7}$ &  $23\pm^{13}_{9}$ &  -  & {\tt C} & \ding{55} &  $14-24$  &  $a$ &                                  324 &  $1.222$  &  $2.32$  &  $7.63$  & $0.8\pm^{0.3}_{0.3}$ &  -  &  -  & {\tt A} & \ding{51} &  $3-11$  &  $a$ \\
84 &  $2.561$  &  $2.44$  &  $88.87$  & $1.7\pm^{0.4}_{0.4}$ &  $14\pm^{8}_{6}$ &  -  & {\tt C} & \ding{55} &  $8.0-23$  &  $a$ &                                                     200 &  $2.567$  &  $10.00$  &  $455.99$  & $1.9\pm^{0.1}_{0.1}$ &  -  &  $0.208\pm^{0.140}_{0.093}$ & {\tt B} & \ding{51} &  $\geq0.94$  &  $-$ &                   326 &  $1.089$  &  $2.80$  &  $17.57$  & $1.9\dagger$ &  -  &  -  & {\tt C} & \ding{51} &  $1.0-1.9$  &  $a$ \\
89 &  $1.271$  &  $5\times10^{-4}$  &  $0.009$  & $1.3\pm^{0.4}_{0.4}$ &  $12\pm^{8}_{5}$ &  -  & {\tt C} & \ding{51} &  $16-26$  &  $a$ &                                          201 &  $2.713$  &  $3.07$  &  $164.74$  & $1.9\pm^{0.1}_{0.1}$ &  -  &  -  & {\tt A} & \ding{51} &  $\geq0.41$  &  $-$ &                                            330 &  $2.208$  &  $5.11$  &  $132.73$  & $1.7\pm^{0.1}_{0.1}$ &  $1.06\pm^{0.80}_{0.74}$ &  -  & {\tt C} & \ding{55} &  $0.8-1.3$  &  $-$ \\
92 &  $3.42$  &  $1.15$  &  $176.05$  & $2.2\pm^{0.4}_{0.4}$ &  $19\pm^{11}_{8}$ &  -  & {\tt C} & \ding{55} &  $11-30$  &  $a$ &                                                    203 &  $0.544$  &  $78.64$  &  $87.31$  & $1.9\pm^{0.0}_{0.0}$ &  -  &  -  & {\tt A} & \ding{51} &  $\geq0.24$  &  $-$ &                                            337 &  $0.837$  &  $25.94$  &  $84.93$  & $1.9\pm^{0.1}_{0.1}$ &  -  &  $0.225\pm^{0.022}_{0.026}$ & {\tt B} & \ding{51} &  $\geq0.24$  &  $-$ \\
93 &  $2.81$  &  $2.48$  &  $122.41$  & $1.7\pm^{0.6}_{0.5}$ &  $19\pm^{14}_{10}$ &  -  & {\tt C} & \ding{51} &  $8.4-33$  &  $a$ &                                                  204 &  $0.599$  &  $4.11$  &  $20.06$  & $2.7\pm^{0.8}_{0.7}$ &  $48\pm^{17}_{13}$ &  -  & {\tt C} & \ding{55} &  $34-65$  &  $a$ &                                 338 &  $1.27$  &  $11.04$  &  $55.82$  & $1.2\pm^{0.3}_{0.2}$ &  $2.68\pm^{1.74}_{1.32}$ &  -  & {\tt C} & \ding{51} &  $1.4-4.4$  &  $a$ \\
95 &  $1.17$  &  $5.29$  &  $71.49$  & $2.1\pm^{0.9}_{0.8}$ &  $65\pm^{36}_{30}$ &  -  & {\tt C} & \ding{55} &  $46-77$  &  $a$ &                                                    210 &  $3.193$  &  $1.22$  &  $115.02$  & $2.0\pm^{0.5}_{0.4}$ &  $7.29\pm^{7.09}_{4.99}$ &  -  & {\tt C} & \ding{55} &  $2-14$  &  $a$ &                           341 &  $1.324$  &  $10.53$  &  $144.78$  & $2.3\pm^{0.0}_{0.0}$ &  -  &  -  & {\tt A} & \ding{51} &  $\geq0.24$  &  $-$ \\
96 &  $0.88$  &  $3.08$  &  $8.56$  & $1.5\pm^{0.2}_{0.1}$ &  -  &  -  & {\tt A} & \ding{55} &  $0.33-0.88$  &  $u$ &                                                                211 &  $2.612$  &  $3.66$  &  $122.21$  & $1.6\pm^{0.2}_{0.2}$ &  $4.32\pm^{2.34}_{1.97}$ &  -  & {\tt C} & \ding{51} &  $2.3-6.7$  &  $a$ &                        344 &  $1.512$  &  $18.92$  &  $206.39$  & $1.6\pm^{0.1}_{0.1}$ &  $10.0\pm^{2}_{2}$ &  -  & {\tt E} & \ding{55} &  $7.1-9.5$  &  $a$ \\
97 &  $2.73$  &  $0.99$  &  $133.75$  & $2.4\pm^{1.2}_{1.0}$ &  $59\pm^{64}_{36}$ &  -  & {\tt C} & \ding{55} &  $24-65$  &  $a$ &                                                   213 &  $2.202$  &  $3.04$  &  $35.26$  & $1.0\pm^{0.2}_{0.2}$ &  -  &  -  & {\tt A} & \ding{55} &  $3.8-7.4$  &  $a$ &                                              345 &  $1.235$  &  $6.91$  &  $69.43$  & $2.2\pm^{0.1}_{0.1}$ &  -  &  -  & {\tt A} & \ding{55} &  $\geq0.24$  &  $-$ \\
98 &  $3.00$  &  $2.38$  &  $119.75$  & $1.7\pm^{0.3}_{0.3}$ &  $5.84\pm^{5.01}_{3.83}$ &  -  & {\tt C} & \ding{51} &  $2.0-11$  &  $a$ &                                            214 &  $1.499$  &  $15.36$  &  $124.88$  & $1.2\pm^{0.2}_{0.2}$ &  $26\pm^{8}_{6}$ &  $0.362\pm^{0.146}_{0.098}$ & {\tt D} & \ding{55} &  $31-47$  &  $a$ &         358 &  $0.976$  &  $68.91$  &  $267.86$  & $1.5\pm^{0.1}_{0.1}$ &  $12\pm^{2}_{1}$ &  $0.278\pm^{0.278}_{0.160}$ & {\tt D} & \ding{51} &  $12-14$  &  $a$ \\
102 &  $1.57$  &  $0.81$  &  $12.55$  & $\leq4$ &  $43\pm^{44}_{6}$  &  -  & {\tt E} & \ding{51} &  $4.8-27$  &  $a$ &                                                               219 &  $2.308$  &  $1.68$  &  $63.03$  & $1.9\pm^{0.2}_{0.2}$ &  -  &  $0.039\pm^{0.155}_{0.009}$  & {\tt B} & \ding{51} &  $\geq0.24$  &  $-$ &                    359 &  $1.574$  &  $34.95$  &  $428.02$  & $1.7\pm^{0.0}_{0.0}$ &  -  &  -  & {\tt A} & \ding{51} &  $\geq0.24$  &  $-$ \\
103 &  $2.03$  &  $9.41$  &  $231.52$  & $1.8\pm^{0.1}_{0.1}$ &  $3.35\pm^{0.79}_{0.73}$ &  -  & {\tt C} & \ding{51} &  $2.6-4.1$  &  $a$ &                                          221 &  $1.220$  &  $2.97$  &  $57.61$  & $2.2\pm^{0.7}_{1.1}$ &  $86\pm^{36}_{48}$ & $0.902\pm^{0.202}_{0.805}$ & {\tt D} & \ding{55} &  $38-79$  &  $a$ &          361 &  $0.977$  &  $20.00$  &  $89.39$  & $1.9\pm^{0.0}_{0.0}$ &  -  &  -  & {\tt A} & \ding{51} &  $\geq0.24$  &  $-$ \\
106 &  $0.667$  &  $2.33$  &  $51.20$  & $1.9\dagger$ & $30\pm^{8}_{8}$  &  -  & {\tt C} & \ding{55} &  $6.6-19$  &  $a$ &                                                            222 &  $0.424$  &  $2.87$  &  $1.93$  & $1.9\dagger$ & $3\pm^{1}_{3}$ &  -  & {\tt C} & \ding{55} &  $0.19-0.46$  &  $u$ &                                          366 &  $2.34$  &  $3.27$  &  $73.65$  & $1.5\pm^{0.2}_{0.2}$ &  -  &  -  & {\tt A} & \ding{51} &  $1.8-3.1$  &  $a$ \\
107 &  $0.665$  &  $6.05$  &  $9.73$  & $1.6\pm^{0.1}_{0.1}$ &  $0.77\pm^{0.22}_{0.20}$ &  -  & {\tt C} & \ding{55} &  $0.57-0.99$  &  $u$ &                                          227 &  $0.834$  &  $8.81$  &  $30.39$  & $1.9\pm^{0.4}_{0.5}$ &  $3.64\pm^{2}_{1}$ & $0.400\pm^{0.200}_{0.310}$ & {\tt D} & \ding{55} &  $1.2-1.5$  &  $a$ &        374 &  $1.041$  &  $13.20$  &  $58.81$  & $1.5\pm^{0.3}_{0.3}$ &  $13\pm^{5}_{3}$ &  -  & {\tt C} & \ding{55} &  $9-18$  &  $a$ \\
108 &  $2.562$  &  $2.99$  &  $109.99$  & $1.7\pm^{0.3}_{0.3}$ &  $30\pm^{12}_{9}$ &  -  & {\tt C} & \ding{51} &  $20-42$  &  $a$ &                                                   228 &  $1.216$  &  $2.84$  &  $21.25$  & $1.8\pm^{0.2}_{0.2}$ &  $2.65\pm^{0.93}_{0.79}$ &  -  & {\tt C} & \ding{51} &  $1.9-3.6$  &  $a$ &                         375 &  $1.350$  &  $4.34$  &  $49.77$  & $2.1\pm^{0.1}_{0.1}$ &  -  & $0.063\pm^{0.052}_{0.087}$ & {\tt B} & \ding{51} &  $---$  &  $-$ \\
109 &  $0.679$  &  $3.48$  &  $29.01$  & $2.7\pm^{1.2}_{1.0}$ &  $71\pm^{34}_{26}$ &  -  & {\tt C} & \ding{55} &  $40-71$  &  $a$ &                                                   230 &  $0.520$  &  $2.16$  &  $2.13$  & $1.8\pm^{0.4}_{0.3}$ &  $1.02\pm^{0.70}_{0.49}$ &  -  & {\tt C} & \ding{55} &  $0.8-1.2$  &  $-$ &                          382 &  $0.527$  &  $46.04$  &  $50.58$  & $1.7\pm^{0.1}_{0.1}$ &  $4\pm^{3}_{4}$ & $0.398\pm^{0.063}_{0.080}$ & {\tt D} & \ding{55} &  $0.18-0.28$  &  $u$ \\
111 &  $1.097$  &  $11.03$  &  $72.33$  & $1.7\pm^{0.2}_{0.2}$ &  $30\pm^{7}_{6}$ &  -  & {\tt E} & \ding{55} &  $23-36$  &  $a$ &                                                    231 &  $1.730$  &  $2.92$  &  $33.05$  & $1.4\pm^{0.4}_{0.4}$ &  $14\pm^{8}_{6}$ &  -  & {\tt C} & \ding{51} &  $7.7-22$  &  $a$ &                                  385 &  $1.220$  &  $7.46$  &  $36.03$  & $1.3\pm^{0.3}_{0.3}$ &  $2.70\pm^{1.59}_{1.23}$ &  -  & {\tt C} & \ding{55} &  $1.5-4.3$  &  $a$ \\
113 &  $0.738$  &  $1.69$  &  $4.20$  & $1.9\pm^{0.6}_{0.5}$ &  $3.60\pm^{2.08}_{1.55}$ &  -  & {\tt C} & \ding{51} &  $2.8-4.6$  &  $a$ &                                            232 &  $1.620$  &  $2.58$  &  $34.80$  & $1.8\pm^{0.1}_{0.1}$ &  -  &  -  & {\tt A} & \ding{55} &  $\leq0.09$  &  $u$ &                                             388 &  $0.605$  &  $4.24$  &  $5.44$  & $1.6\pm^{0.2}_{0.1}$ &  -  &  -  & {\tt A} & \ding{55} &  $\geq0.24$  &  $-$ \\
114 &  $1.806$  &  $5.54$  &  $15.74$  & $\leq5$ &  -  &  -  & {\tt A} & \ding{51} &  $22-120$  &  $a$ &                                                                              233 &  $1.615$  &  $5.67$  &  $75.77$  & $1.7\pm^{0.4}_{0.3}$ &  $12\pm^{5}_{4}$ &  $0.116\pm^{0.087}_{0.055}$ & {\tt D} & \ding{51} & $14-32$ & $a$ &              390 &  $1.118$  &  $6.30$  &  $41.36$  & $1.9\pm^{0.1}_{0.1}$ &  -  &  -  & {\tt A} & \ding{55} &  $\geq0.24$  &  $-$ \\ \hline\hline
\end{tabular}}}
\end{minipage}}
\end{table*}

\begin{table}
    \caption{Summary of the mean value of the general X-ray properties
      given by the best fitted model (see Section \ref{section3}). The
      last column is the probability that both distributions (IR
      power-law and IR non-power-law sample) come from the same
      distribution according to the Kolmogorov-Smirnov test.}
    \label{tab:fittedParameters}
    \centering
    \resizebox{0.47\textwidth}{!}{ %
    \begin{tabular}{|c|ccc||c|}\hline
        {\bf X-ray parameter}                   & {\bf Sample}       & {\bf IR pl}         & {\bf IR no-pl}   & {\bf KS} \\\hline\hline
        $\langle\Gamma\rangle$                  & $1.72\pm0.36$      & $1.74\pm0.36$       & $1.71\pm0.37$    & $\sim65\%$     \\  
        $\langle\log(N_{H}^{intr})\rangle$      & $22.00\pm1.81$     & $22.13\pm1.36$      & $21.89\pm1.34$   & $\lesssim10\%$ \\  
        $\langle z\rangle$                      & $1.42\pm0.81$      & $1.81\pm0.79$       & $1.15\pm0.70$    & -- \\
        $\langle\log(F_{2-10\,keV})\rangle$     & $-14.35\pm0.63$    & $-14.33\pm0.91$     & $-14.36\pm0.32$  & $\sim 30\%$    \\  
        $\langle\log(L_{2-10\,keV})\rangle$     & $43.50\pm0.84$     & $43.80\pm0.99$      & $43.28\pm0.63$   & $\lesssim3\%$ \\\hline 
    \end{tabular}
    }
\end{table}

\begin{table}
    \centering
	\caption{Summary of the results of the spectral fitting.}
	\label{tab:BestFittedModelSummary}
    \resizebox{0.45\textwidth}{!}{
    \begin{tabular}{|llr|cc|cc|cc|}\hline
        & & Model  & 
        \multicolumn{2}{c|}{N} & \multicolumn{2}{c|}{$N_{IR pl}$} & \multicolumn{2}{c|}{$N_{IR non-pl}$}  \\
        & & (1) & \multicolumn{2}{c|}{(2)} & \multicolumn{2}{c|}{(3)} & \multicolumn{2}{c|}{(4)} \\\hline\hline
        & & {\tt A}: \emph{powerlaw}                         & 45 & ($31\pm^{6}_{6}$)  & 15 &($25\pm^{10}_{8}$)   &  30 &($34\pm^{9}_{8}$) \\ 
        & & {\tt B}: \emph{zbbody} $+$ \emph{powerlaw}       & 13 & ($9\pm^{4}_{3}$)   & 9  &($15\pm^{8}_{6}$)    &  4  &($5\pm^{4}_{3}$)  \\
        & & {\tt C}: \emph{zphabs} $\times$ \emph{powerlaw}  & 69 & ($47\pm^{7}_{7}$)  & 27 &($45\pm^{10}_{-10}$)  &  42 &($48\pm^{9}_{8}$)  \\
        & & {\tt D}: \emph{zbbody} $+$ (\emph{zphabs} $\times$ \emph{powerlaw})  & 12 & ($8\pm^{4}_{3}$) & 5  &($8\pm^{7}_{5}$) & 7 &($8\pm^{6}_{4}$) \\
        & & {\tt E}: \emph{powerlaw} + (\emph{zphabs} $\times$ \emph{powerlaw})  & 8  & ($5\pm^{4}_{3}$) & 4  &($6\pm^{7}_{4}$) & 4 & ($5\pm^{4}_{3}$) \\\hline
    \multicolumn{9}{l}{Notes.}\\
    \multicolumn{9}{l}{(1) XSPEC model definition: \emph{power-law} as a simple photon power-law; \emph{zphabs} as a rest-} \\
    \multicolumn{9}{l}{frame photoelectric absorption; \emph{zbbody} as a redshifted blackbody spectrum.}\\
    \multicolumn{9}{l}{All models include Galactic absorption (\emph{phabs}).}\\
    \multicolumn{9}{l}{(2) Total number and fraction of best fitted AGN with the indicated model.}\\
    \multicolumn{9}{l}{(3) Number and fraction of IR \emph{power-law objects} best fitted with the indicated model.}\\
    \multicolumn{9}{l}{(4) Number and fraction of IR \emph{non-power-law objects} best fitted with the indicated model.}\\
    \end{tabular}
    }
\end{table}

\begin{itemize}
    \item \emph{Soft X-ray component}. 

 Among the whole sample, only $33$ sources ($\sim$$22^{+6}_{-2}$\%)
 have a soft emission component ({\tt B}, {\tt D}, and {\tt E}
 models). In the full X-ray spectral analysis presented in
 \citet{Comastri2013} the origin of this soft X-ray emission component
 is studied in more detail.  We note that the low fraction of sources
 with a significant soft X-ray component may be due to the high
 redshift of our sources (see Figure~\ref{fig:sample_properties}), for
 which the soft component is often shifted below the used X-ray
 spectral band.\\

\begin{table}
    \caption{Summary of the sources in our sample that were identified as Compton-Thick candidates (CThick) in other works:
    \emph{us}: this work; \emph{T06}: \citet{Tozzi2006}; \emph{C11}: \citet{Comastri2011}; \emph{B12}: \citet{Brightman2012};
    \emph{I12}: \citet{Iwasawa2012}; and \emph{G13}: \citet{Georgantopoulos2013}. 
    }
    \label{tab:CT}
    \centering
    \resizebox{0.49\textwidth}{!}{ %
    \begin{tabular}{|c|cccccc|}\hline
        {\bf ID210} & us & T06 & C11 & B12 & I12 & G13 \\\hline
        30  & heavily    & -  & -  & -     & heavily  & -             \\
        44  & unabsorbed & CThick& -  & heavily & -       & -             \\
        48  & moderately & -  & -  & -     & -       & heavily   \\
        64  & heavily    & -  & -  & -     & heavily  & -             \\
        66  & heavily    & -  & -  & -     & -       & secure-CThick     \\
        106 & moderately & CThick& -  & heavily & -       & -             \\
        114 & heavily    & -  & -  & -     & heavily & -             \\
        144 & heavily    & CThick& CThick& heavily & heavily & secure-CThick     \\
        147 & heavily    & CThick& CThick& heavily & -       & secure-CThick     \\
        155 & moderately & CThick& -  & heavily & -       & -             \\
        180 & heavily    & -  & -  & -     & heavily & heavily   \\
        214 & heavily    & -  & -  & -     & -       & heavily   \\
        222 & unabsorbed & -  & -  & -     & -       & heavily   \\
        245 & heavily    & -  & -  & -     & heavily & heavily   \\
        289 & moderately & -  & -  & -     & -       & heavily   \\
        324 & unabsorbed & -  & -  & -     & -       & secure-CThick     \\\hline
    \end{tabular}
    }
\end{table}

    \item \emph{Compton-Thick candidates}.

The simple power-law model gives a good fit for $58$
($39^{+6}_{-7}$\%) X-ray sources, i.e. {\tt A} and {\tt B} models.
The remaining $89$ ($61^{+6}_{-7}$\%) sources display a best fit with
X-ray obscuration, i.e., {\tt C}, {\tt D}, and {\tt E} models.  The
intrinsic absorption distribution measured is bimodal (see
Figure~\ref{fig:fittedParameters}), in the sense that $39^{+7}_{-6}\%$
of sources have very small $N_{H}^{intr}$ (below the Galactic value)
and appear separated from the distribution of the bulk of the sources.
The lowest bin in the distribution (i.e., that with
$N_{H}^{intr}\simeq10^{20}$ cm$^{-2}$) accumulates those sources for
which we cannot measure any intrinsic absorption, i.e. it includes all
the sources best fitted with an unabsorbed model (model {\tt A} and
{\tt B}, see Table~\ref{tab:BestFittedModelSummary}). The last bin at
$N_{H}^{intr}=10^{24}$ cm$^{-2}$ includes the few sources with
$N_{H}^{intr} >10^{24}$ cm$^{-2}$ but their error bars make them
compatible with being \emph{Compton-Thin}. Thus, there are no
\emph{Compton-Thick} sources securely detected in our sample using the
first-order spectral model adopted in this paper (see
Table~\ref{tab:CT} and further discussion below).  In
Figure~\ref{fig:fittedParameters}, we also show the normalised
distribution of the intrinsic absorption for the sample by
\cite{Tozzi2006}.  The excess of sources in the lowest bin of our
  sample can be explained through the lack of sources with detected intrinsic
  absorption to $N_H^{intr}\sim10^{21-22}$ cm$^{-2}$, as we are
  really not sensitive for most of our sources to intrinsic column
  densities to $N_H^{intr}\sim10^{21-22}$ cm$^{-2}$, especially at high redshifts. We remark that both
distributions have in general a similar shape but they differ at large
absorptions (see Figure~\ref{fig:fittedParameters}, bottom left
panel), in the Compton-Thick region. \\
In summary, $69$ objects appear to be unabsorbed X-ray sources, i.e.,
their best-fit X-ray absorption column is $N_{H}^{intr}<10^{22}$
cm$^{-2}$. That is, $78\,(53^{+7}_{-7}\%)$ of the galaxies in our
sample host significantly obscured active nuclei what is not far but
still on the lower side from that predicted by XRB synthesis models
($\sim 64\%$, \citealt{Gilli2007}). $60\,(77^{+16}_{-11}\%)$ of 78 are
Compton-Thin with a moderate obscuration ($N_{H}^{intr}\leq
3\times10^{23}$ cm$^{-2}$) and the remaining $18$ objects are heavily
obscured AGN. Among these, we find two Compton-Thick candidates in the
sample (right-most bin in the bottom panel in
Figure~\ref{fig:fittedParameters}), in the sense that the column
density has some likelihood of exceeding $1.5 \times 10^{24}\, {\rm
  cm}^{-2}$.  However, the uncertainty in the intrinsic column
densities of these sources (ID210=66 IR non-power-law, and ID210=144
IR power-law), is such that their Compton-Thin nature cannot be ruled
out. If additional criteria are introduced this dichotomy could be
resolved and more sources could be identified as Compton-Thick. \\

For example, we find that our best-fit value of the column
  density to sources ID210= 144, 147 using an absorbed power-law model
  is in very good agreement with the fit to the \xmm{} spectrum of the
  same sources with the more sophisticated torus model of Murphy \&
  Yaqoob (2009) by \cite{Comastri2011} using the same ultra-deep \xmm{}
  data in the CDFS. The uncertainties on the fitted column density do
  not allow us to assert that they are Compton-Thick, but
  \citet{Comastri2011} use the presence of a strong Iron line and 
  a strong reflection component to classify source ID210 = 147 as such.

 Similarly, \citet{Georgantopoulos2013} have conducted a search for
 Compton-Thick AGN using the same \xmm~ exposures. The spectral model
 that they fitted to the data is \textrm{PLCABS} \citep{Yaqoob1997},
 which is specially tailored to Compton-Thick sources, as it properly
 takes Compton scattering and reflection up to column densities up to
 $\sim 5\times 10^{24}\, {\rm cm}^{-2}$ into account. A total of 9
 candidate Compton-Thick sources (numbers ID210=48, 66, 144, 147, 214,
 222, 245, 289 and 324) were found by them. Our analysis of the X-ray
 spectra of these 9 sources is coincident with theirs in terms of
 best-fit spectral index and absorbing column.  They classify four of
 these (ID210=144, 66, 147, 324) as Compton-Thick: the first two had
 transmission-dominated spectra with strong Iron lines (we also find
 them to be heavily absorbed) and the last two because of their
 reflection-dominated spectra (we also detect 66 as a strongly
 absorbed source, but not 324, because in our fit a flat unabsorbed
 model mimics its intrinsic reflection-dominated spectrum).

 Another paper using the same \xmm{} exposures dealing with
  obscured AGN is \citet{Iwasawa2012}, who label sources ID210= 30,
  64, 144, 180, 245, 114 as strongly absorbed sources in agreement
  with our findings. Again, the last source was found by them to be a possible
  Compton-Thick candidate because of its reflection-dominated
  spectrum, while our best-fit model is a flat spectrum with a
  moderate column density.

 Other works to characterize obscuration using deep X-ray data in
  the same region of the sky have also been carried out by
  \citet{Tozzi2006} and \citet{Brightman2012}. The sources ID210=44,
  106, 155, 144, and 147 were identified as Compton-Thick sources by
  \cite{Tozzi2006} with their $1$ Ms \emph{Chandra} data. We disagree
  with the Compton-Thick character of ID210=44, 106, and 155, which
  appear to be at most moderately absorbed in our analysis, while the
  last two would be heavily absorbed, in agreement with
  \cite{Brightman2012}, but see the discussion above about those last two sources using additional criteria.

While all models predict that Compton-Thick sources have an important
role to play in filling the so far unresolved XRB above $\sim 8\, {\rm
  keV}$, such sources have been generally elusive in X-ray surveys
conducted both with \chandra~ and \xmm. This is also evidenced in the
current work, which shows that even with the deepest exposures it is
very difficult to unambiguously find sources with column densities in
excess of $10^{24}\, {\rm cm}^{-2}$. In the case of our \xmm~ data
this is likely due to a combination of large background at high
energies together with a modest (but still significant) effective
area.  In the case of \chandra~ data, a much more reduced background
(since the better angular resolution enables a much smaller extraction
region for the X-ray spectra) is hardly compensated by a smaller
effective area with respect to \xmm. Detecting significant numbers of
Compton-Thick sources remains a challenge with current X-ray
instruments. \\
 

\item \emph{IR power-law versus IR non-power-law from best-fit model}.
  On the one hand, we found that both the measured X-ray photon index
  and the observed X-ray flux remain virtually confined to the mean
  value of $\langle\Gamma\rangle=1.72\pm0.36$ and $\langle
  \log{(F_{2-10\,keV})}\rangle = -14.35\pm 0.63$ erg/s/cm$^{2}$,
  respectively. There is one unabsorbed AGN (ID210=114 IR power-law)
  for which the resulting photon index is $>3$.  This very steep slope
  is probably a consequence of both the small number of counts and
  high background in its X-ray spectrum, although we cannot exclude a
  very high accretion rate source, e.g., a Narrow-Line Seyfert 1
  galaxy. \\ On the other hand, the distribution of the intrinsic
  absorption and the rest-frame 2-10 keV luminosity appears to span
  distinct regions (see Section~\ref{intrinsicabsorption}). There are
  some evidence that the distribution of the intrinsic absorption
  column densities seems to have a different shape for IR power-law
  and IR non-power-law populations according to the Kolmogorov-Smirnov
  test ($\lesssim10\%$), which suggests that both samples might have
  different parent distributions. Finally, we also found that the
  rest-frame 2-10 keV luminosity distribution appears to have a
  different shape for IR power-law and IR non-power-law populations
  according to the Kolmogorov-Smirnov test ($\lesssim3\%$). The mean
  values for the logarithmic X-ray luminosity are $43.8\pm0.99$ and
  $43.28\pm0.63$ for the IR power-law and IR non-power-law
  populations, respectively. The IR power-law population appears to
  select better higher luminosity AGN because as expected, the IRAC
  selection cannot efficiently identify low-luminosity AGN, and
  appears to be incomplete for low luminosity AGN (see
  \citealt{Donley2007,Donley2008}).

\end{itemize}


\section{Intrinsic absorption}
\label{intrinsicabsorption}

To explore and quantify the efficiency in finding
highly-obscured AGN and/or QSO when selecting the sources as IR
power-law and IR non-power-law, all individual spectra were re-fitted
by an absorbed power-law model (plus a soft-excess component when
required by the X-ray data), i.e., using models {\tt C}, {\tt D} or
{\tt E}.  In addition, now, for those sources with a small number of
counts ($<500$), the X-ray power-law photon index ($\Gamma$) was fixed
to $1.9$ (the typical average value for unabsorbed AGN) to constrain
better the intrinsic absorption. As expected, for the sources,
  which had already been fitted with models {\tt C}, {\tt D}, or {\tt
    E} the results remain essentially unchanged, the only difference
  is that now we assign a $1\sigma$ uncertainty interval to the
  intrinsic column density. Similarly, for the sources whose best-fit
  model was {\tt A} or {\tt B} the results are essentially compatible
  except for 8 sources, which are now classified as absorbed. This
  happens because of two reasons: on two cases, the photon index is
  now fixed to 1.9, because of the low number of counts, hence the
  fitted absorption increased considerably because these sources had
  low values of the photon index; on the other six sources, the F-test
  probability of models {\tt C},{\tt D},{\tt E} versus {\tt A},{\tt B}
  was $\gtrsim$90\% (hence the former were not significantly
  better than the latter, according to our criterion in Section
  \ref{section3}: the 95\% confidence interval on $N_H^{intr}$
  includes zero) while the bottom of the 1$\sigma$ confidence interval
  on $N_H^{intr}$ is above $10\times22$~cm$^{-2}$ (i.e., they are
  absorbed according to our classification in this Section). The
  intrinsic hard X-ray luminosities derived in this Section and in
  Section~\ref{section3} are very similar, solely in two cases the
  ratio between both luminosities exceeds a factor 1.5, but in all
  cases the source remains in the same luminosity bin (defined
  below). 

\begin{itemize}
   
    \item \emph{Absorbed fraction from best-fit values}.

      We have subdivided our IR power-law and IR non-power-law samples
      according to their $N_{H}^{intr}$ values, also taking the
      1$\sigma$ uncertainty intervals into account, in the sense that
      those sources whose 1$\sigma$ uncertainty interval is fully
      above (below) $10^{22}$ cm$^{-2}$ are called absorbed
      (unabsorbed) at the 1$\sigma$ level.  Those objects whose
      1$\sigma$ interval crosses the $10^{22}$ cm$^{-2}$ border have
      been labelled as \emph{unclassified}.  The number of sources in
      each of these subdivisions in bins of $2-10$ keV luminosity is
      listed in Table~\ref{tab:absorbedfractions} and the individual
      classification of each source is shown in
      Table~\ref{tab:Properties_of_the_sample}.  As a further test of
      the robustness of our estimates, we also redefined the absorbed
      and unabsorbed samples using a 2$\sigma$ threshold around
      $10^{22}\, {\rm cm}^{-2}$. The number of sources in each
      subsample change very little, and our conclusions remain
      unaffected. In what follows, in our conservative approach those
      sources whose were classified as \emph{unclassified} will be
      considered as unabsorbed.\\

\begin{table*}
    \caption{Number of sources classified as absorbed/unabsorbed (at
      1$\sigma$ significance) and unclassified as a function of the
      intrinsic 2-10 keV luminosity according to their intrinsic
      column density. The $N_{H}^{intr}$ is obtained assuming an
      absorbed power-law model (i.e., {\tt C}, {\tt D}, and {\tt E}
      model). We remark that absorbed/unabsorbed and unclassified
      groups are disjoint sets.}
    \label{tab:absorbedfractions}
    \centering
    \resizebox{0.8\textwidth}{!}{ %
    \begin{tabular}{|c|cccc|cccc|cccc||c|}\hline

    &\multicolumn{4}{c|}{{\bf Absorbed}}&\multicolumn{4}{c|}{{\bf Unclassified}} & \multicolumn{4}{c||}{{\bf Unabsorbed}}&
     \\ \cline{2-13}
    {\bf Sample} & \multicolumn{4}{c|}{$\log L_{X}$ [erg/s]} & \multicolumn{4}{c|}{$\log L_{X}$ [erg/s]}& 
    \multicolumn{4}{c||}{$\log L_{X}$ [erg/s]} &  {\bf Total}\\
    & $<43$ & $43-44$ & $\geq44$ & all & $<43$ & $43-44$ & $\geq44$ & all & $<43$ & $43-44$ & $\geq44$ & all & \\\hline\hline
    {\bf IR power-law }    & 5 & 15 & 21 & 41 &0  & 6  & 7 & 13 &1  & 0  & 5 & 6  & 60\\ \hline
    {\bf IR non-power-law} & 5 & 28 & 9  & 42 &10 & 10 & 4 & 24 &10 & 10 & 1 & 21 & 87\\ \hline\hline
    {\bf Ttotal}              & 10& 43 & 30 & 83 &10 & 16 & 11& 37 &11 & 10 & 6 & 27 & 147\\\hline
    \end{tabular}
    }
\end{table*}

We measured a significant intrinsic absorption in excess of $10^{22}\,
{\rm cm}^{-2}$ for $83$ sources {\bf (see Table~\ref{tab:Properties_of_the_sample})}, of which 41 are IR power-law and 42
are IR non-power-law AGN. A further 27 galaxies are classified as
unabsorbed, out of which 6 are IR power-law and 24 IR non-power-law
galaxies. 37 sources were unclassified at 1$\sigma$ level. \\

An important result from this work is that the fraction of absorbed
sources (see Table~\ref{tab:absorbedfractions}) is higher among the IR
power-law galaxies (41 out of 60) than among the IR non-power-law
galaxies (42 out of 87). Binomial error estimates return a fraction of
absorbed sources among the IR power-law galaxies of $68_{-10}^{+9}$\%
and of $48_{-8}^{+9}$\% for the IR non-power-law galaxies. A Bayesian
estimate of the probability of these two fractions coming from the
same parent distribution \citep{Stevens2005} yields a probability of
0.0023, and therefore the fraction of absorbed sources among IR
power-law galaxies is higher than that of IR non-power-law galaxies at
about the 3$\sigma$ level.\\

    \item \emph{Dependence on X-ray luminosity from best-fit values}.

           Our next step was to check for a possible dependence of the
           column density on luminosity, within the two subsamples.
           We selected three luminosity ranges, $\leq10^{43}\, {\rm
             erg}\, {\rm s}^{-1}$, $10^{43}-10^{44}\, {\rm erg}\, {\rm
             s}^{-1}$, and $>10^{44}\, {\rm erg}\, {\rm s}^{-1}$ with
           median redshifts of $0.65\pm0.11$, $1.59\pm0.69$, and
           $2.21\pm0.75$, respectively. We compute the fraction of IR
           power-law (IR non-power-law) sources, which have been
           classified as \emph{absorbed} AGN at each luminosity range,
           understood as the ratio of the number of absorbed IR
           power-law (IR non-powerlaw) at 1$\sigma$ level to the total
           number of IR power-law (IR non-powerlaw) sources at this
           luminosity bin. We found that the fraction of absorbed
           sources appears roughly constant with luminosity for the IR
           power-law AGN, while it grows from $\sim20\%$ to $\sim64\%$
           for the IR non-power-law galaxies (see
           Figure~\ref{fig:fabs_Lxbins}). \\

           \begin{figure}
             \centering
             \includegraphics[width=0.49\textwidth]{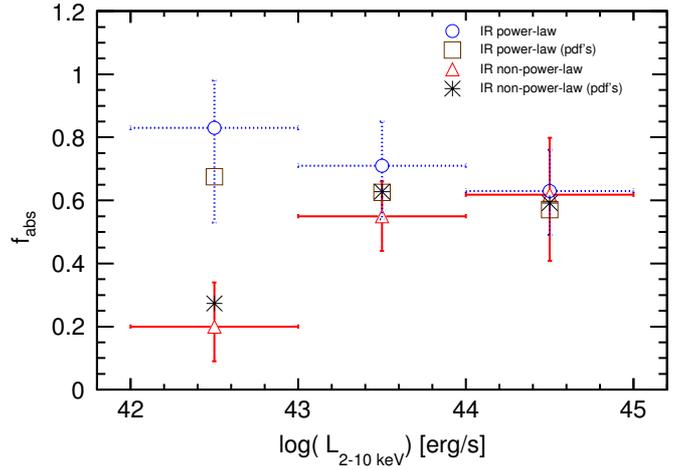}
             \caption{Fraction of absorbed sources (at $\geq1\sigma$) for IR
               power-law and IR non-power-law galaxies as a function of the
               rest-frame absorption corrected 2-10 keV luminosity. The open
               squares and star symbols show the fraction of absorbed sources
               taking into account the probability density function for
               $N_H^{intr}$ of each individual source for IR power-law and IR
               non-power-law, respectively (see Section 5). Note that
               these fractions refer exclusively to our sample and not to the
               overall AGN population.} 
             \label{fig:fabs_Lxbins}
           \end{figure}

    \item \emph{Dependence on X-ray luminosity from probability density functions}.

      \begin{figure*}
        \centering
        \includegraphics[width=0.48\textwidth]{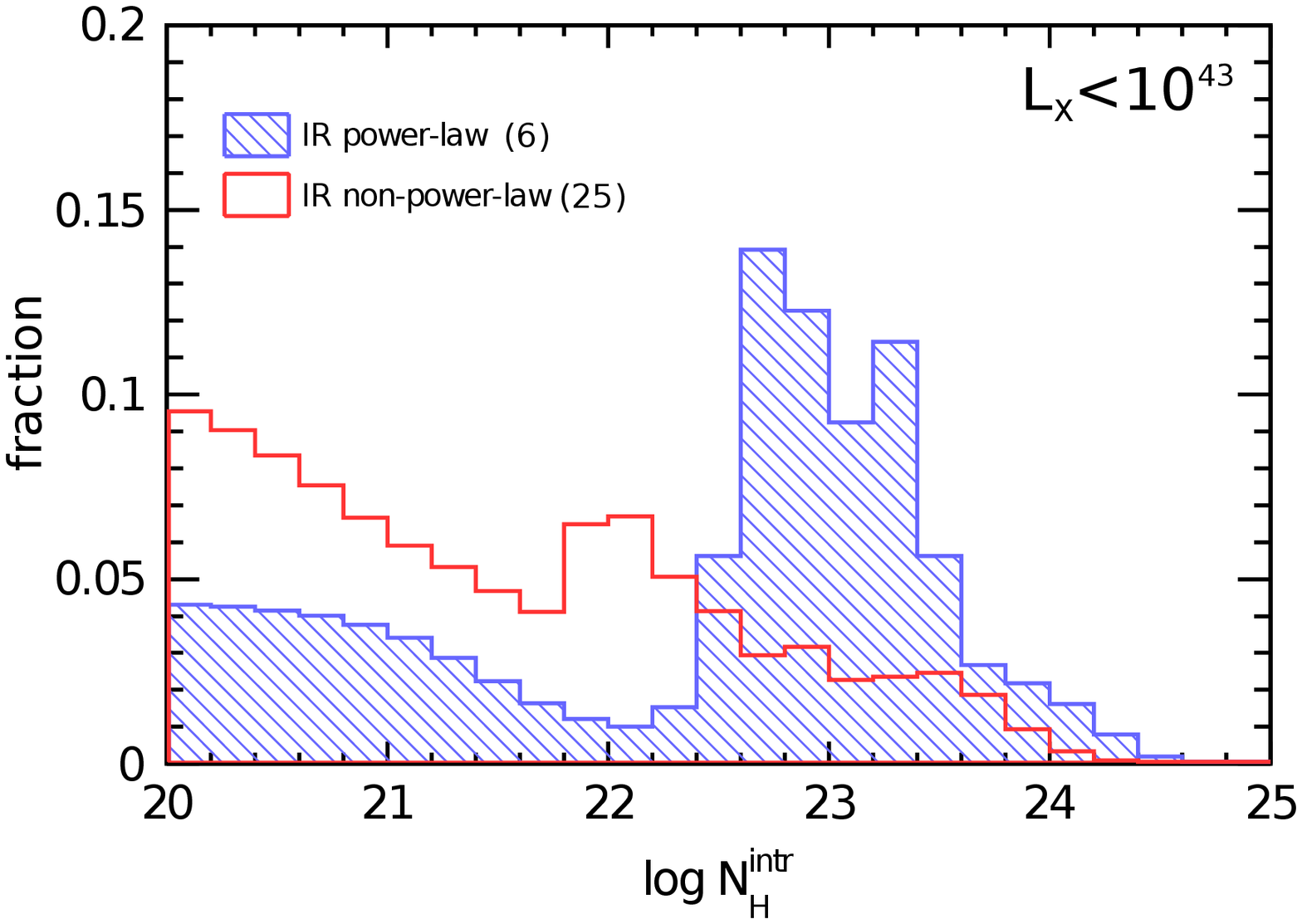}
        \includegraphics[width=0.48\textwidth]{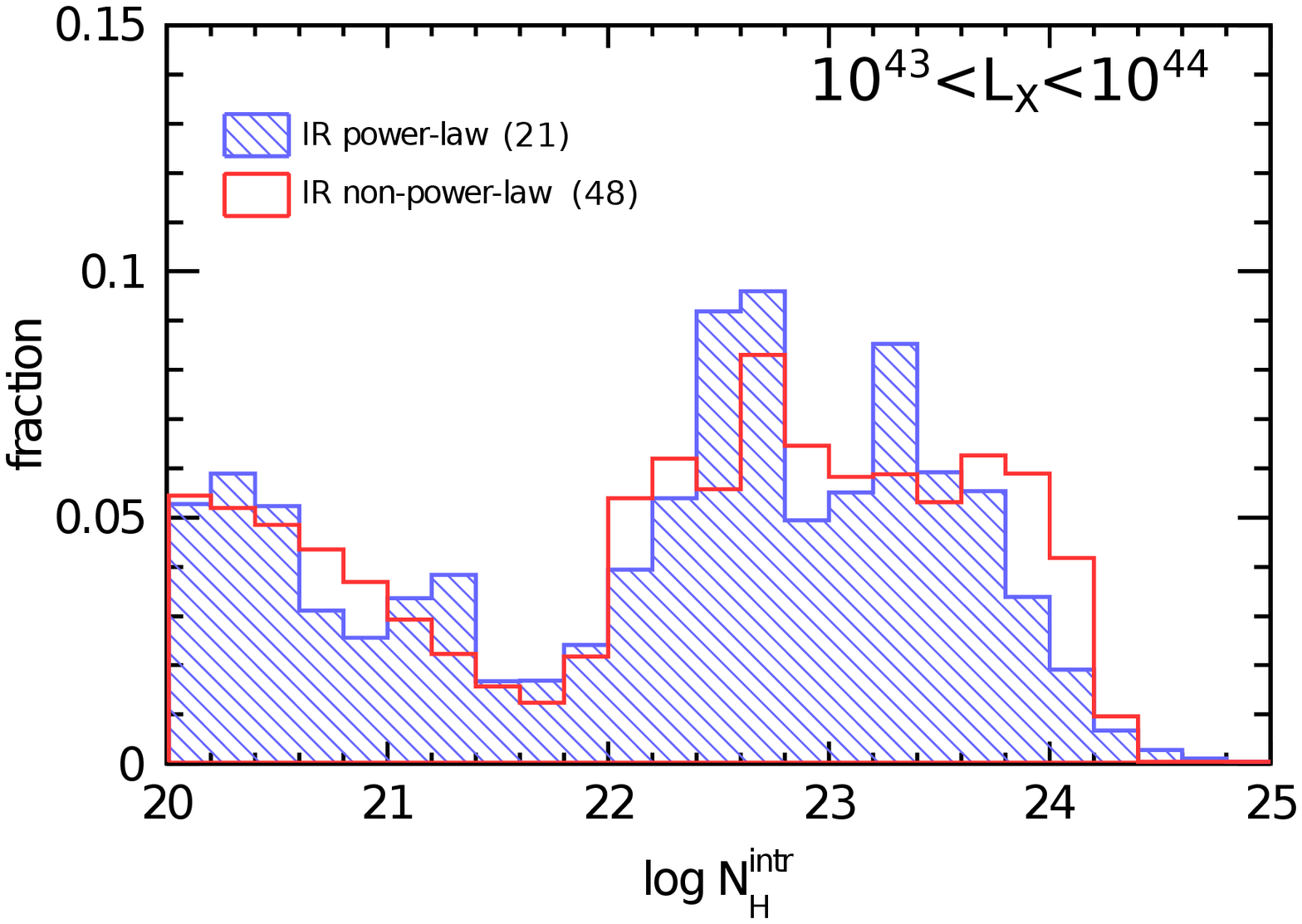}\\
        \includegraphics[width=0.48\textwidth]{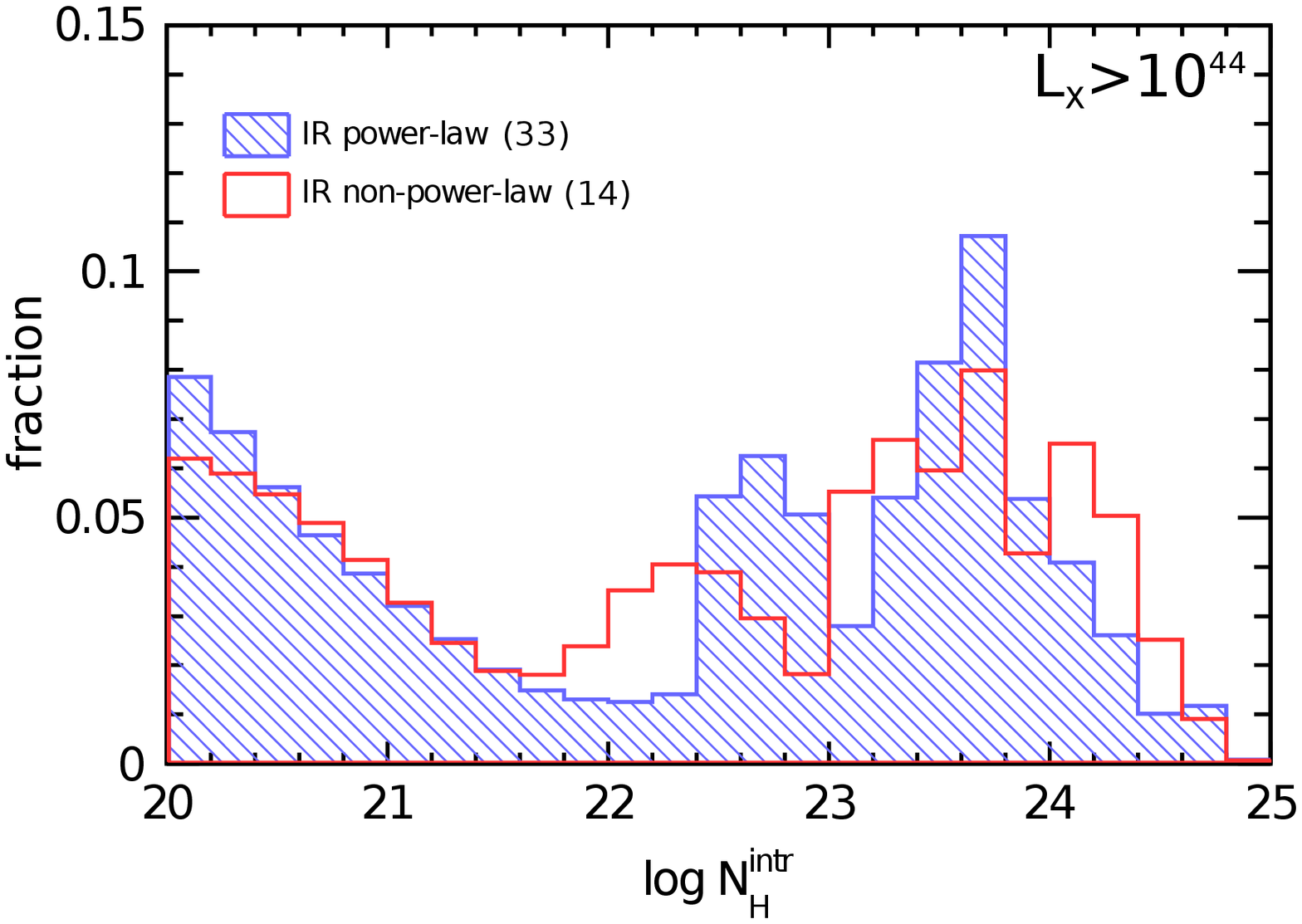}
        \includegraphics[width=0.48\textwidth]{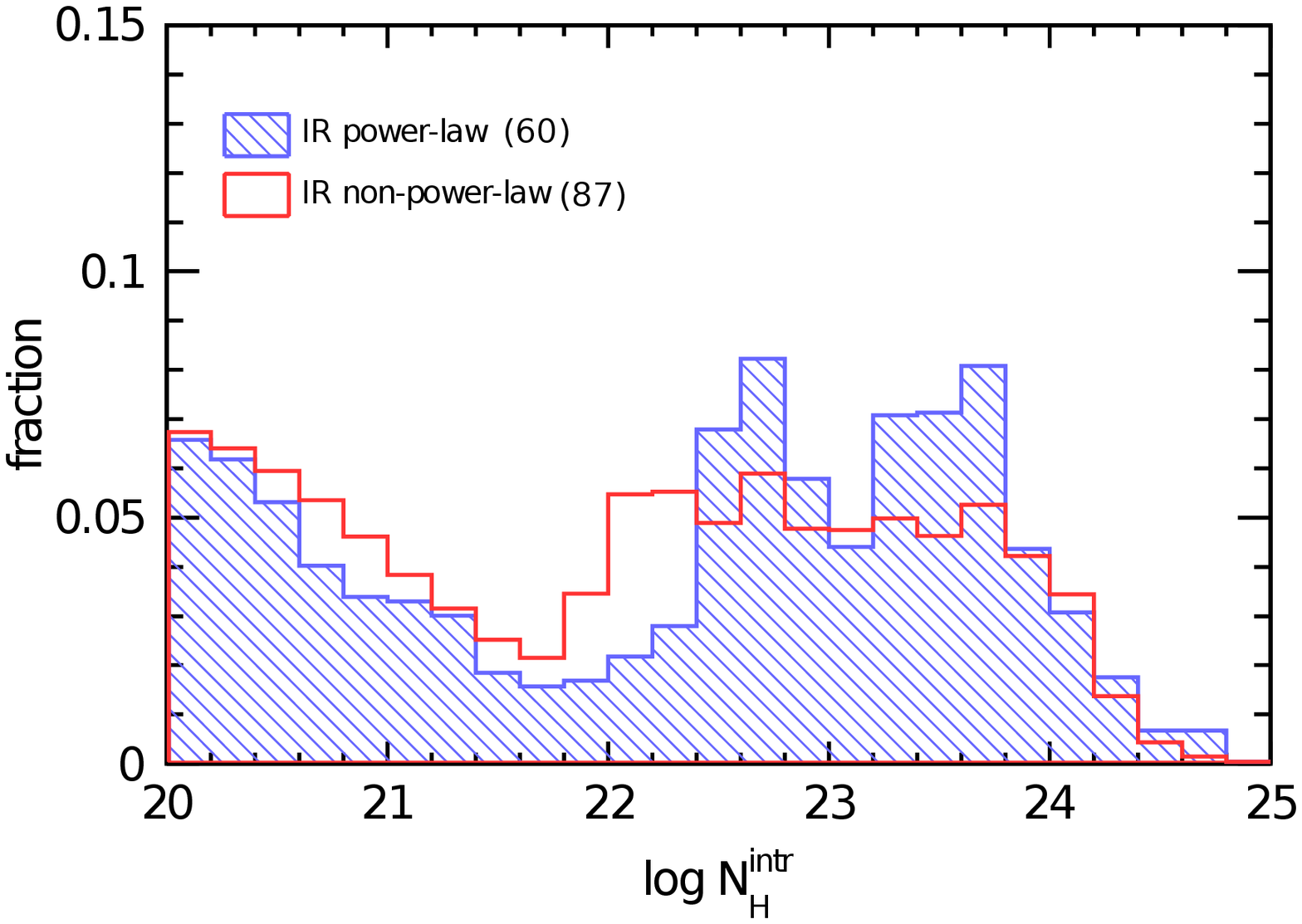}
        \caption{Normalized distribution of intrinsic absorbing column
          densities (in log units) in three absorption corrected 2-10 keV
          luminosity bins (upper panels and lower left panel) and for the
          entire sample (lower right panel). The distributions were
          computed using the probability density function for $N_H^{intr}$
          of each individual source (see Section 5, for details).}
        \label{fig:Nhz_pdfs}
      \end{figure*}

      To take the full distribution of probabilities when building the
      $N_H^{intr}$ distributions into account, we followed the
      following procedure. First, when fitting the X-ray spectrum of
      each individual source, we used the {\tt steppar} command under
      the {\tt XSPEC} fitting package, to find out how the $\chi^2$
      varies with $\log(N_H^{intr})$.  We explored sufficiently wide
      ranges of $\log (N_H^{intr})$ in such a way that
      $\Delta\chi^2=\chi^2-\chi^2_{min}$ (where $\chi^2_{min}$
      corresponds to the best-fit $N_H^{intr}$ value) reaches
      sufficiently high values (hence low probabilities, see
      below). We set an absolute minimum of $\log (N_H^{int})=20.0$,
      because, as discussed before, our data are not sensitive to
      lower values of the intrinsic absorption. Next, we constructed a
      probability density function for $N_H^{intr}$ of each individual
      source by using
      $p(\log(N_H^{intr}))\propto\exp(-\Delta\chi^2/2)$, and
      normalised it between $\log (N_H^{int}/{\rm cm}^{-2})=$20 and
      25. We registered these $p(\log(N_H^{intr}))$ into a common grid
      of values of $\log (N_H^{intr})$ spaced 0.2 units.
      of X-ray luminosities and for the full sample, also taking both
      IR power-law and IR non-power-law populations into account,
      normalising each of the summed probabilities to a total unit
      area (see Figure~\ref{fig:Nhz_pdfs}).\\
      The normalised distributions of $N_H^{intr}$ for the entire sample
      show some difference between the two populations of IR power-law and
      IR non-power-law galaxies, with the former being more heavily absorbed
      than the latter. This behaviour is not evident at high or intermediate
      luminosities, where both populations have an overall indistinguishable
      distribution. However, at the lowest luminosity regime, the
      distributions appear to have a quite different shape. Whilst the
      $N_H^{intr}$ distribution for the IR power-law sample peaks at $\sim
      10^{23}$ cm$^{-2}$, the IR non-power-law sample appears to have
      largely unabsorbed sources. This is somewhat unexpected because of the
      known incompleteness of the power-law selection at low AGN
      luminosities.  We interpret this as due to the different shapes of
      the IR AGN SEDs of type-1 and type-2, with type-2 having steeper
      SEDs \citep[see ][Fig. 9]{RamosAlmeida2011}. The host galaxy, on the
      other hand, has a ``broad'' bump peaking at $\sim$1.6$\mu$m
      (rest-frame). Therefore, the combination of a relatively low
      luminosity type-1 AGN and its host galaxy would result in a
      relatively flat spectral shape in the IRAC bands, likely not meeting
      the IRAC IR power-law criteria.\\ As a further test of the
      robustness of our estimates, we also measured the fraction of
      absorbed sources in both samples at each luminosity bin using the
      probability density function for $N_H^{intr}$ of each individual
      source. The absorbed source fractions change very little (see
      Figure~\ref{fig:fabs_Lxbins}) both results being
      compatible. Moreover, if we compute the absorbed source fraction
      without those sources with photometric redshift (11/60 IR power-law
      and 12/87 IR non-power-law sources), our findings are also in good
      agreement when accounting for uncertainties.\\

\end{itemize}

Therefore, taking the full probability distribution of
$\log(N_H^{intr})$ into account, our analysis is in full agreement with the one
using just the 1$\sigma$ confidence levels: the fraction of absorbed
sources among IR-power-law AGN is higher both in the full sample and
among lower luminosity sources ($\log L_X<43\,{\rm erg/s}$), while it
is compatible with being similar at higher luminosities.  We note
  that, on the one hand, we are not trying to determine the absolute
  fraction of absorbed sources as a function of the X-ray luminosity,
  but to evaluate the effectiveness of the mid-IR selection to identify
  obscured AGN by comparing the fraction of absorbed sources inside
  and outside the IRAC-selection wedge. On the other hand, we do not
  apply any completeness correction to the absorbed source fractions,
  thus these fractions of absorbed sources should not be compared with
  results referring to the overall AGN population
  \citep{Ueda2003,Treister2005,Gilli2007}. We also point out that the
  difference on the fraction of obscured AGN at the lowest luminosity
  bin should be interpreted with caution, since there are only 6 IR
  power-law sources in this low-luminosity bin.\\

 In conclusion, although the IR power-law selection only picks up
  60/147 (about 41\%) of our X-ray-selected high-spectral-quality
  sources, at high X-ray luminosities it singles out about 70\% of our
  sources (33/47). Concentrating on the IR power-law sources, the
  overall percentage of such sources, which are absorbed is 68\%
  (41/60), essentially independent of the AGN luminosity, better than
  the overall fraction of absorbed sources in our full sample (83/147
  $\sim$56\%) and significantly higher than that of IR
  non-power-law sources (42/87$\sim$48\%).  As \citet{Donley2012}
  found, the IR power-law selection produces an incomplete census of
  AGN, its completeness being a strong function of AGN luminosity. Our
  overall estimate of its efficiency to find absorbed sources
  ($\sim$70\%) is similar, if somewhat lower, than their estimate of
  $\sim$75\%.

%
%
%

\section{Summary and conclusions}

In the paper we have investigated the subset of X-ray sources in the
ultra-deep \xmm~ of the Chandra Deep Field South \citep{Ranalli2013}
with a highly significant ($>$8$\sigma$) detection, high exposure time
($>$1Ms) and known (spectroscopic or photometric) redshift, totalling
147 sources.  All of them turn out to have Spitzer/IRAC counterparts,
and they are all detected in the four IRAC bands. Consequently,
  our final sample is biased towards high signal-to-noise X-ray
  sources and cannot be considered as a complete sample of X-ray
  selected or mid-IR-selected sample. 
  However, this does not affect the work main goal of this: to test
  the efficiency of the IR power-law in selecting absorbed X-ray
  sources. We have used the IRAC photometry to classify these
sources into IR power-law and IR non-power-law galaxies, according to
whether or not their IR SED has a power-law-like shape and is
monotonically increasing, following \citet{Donley2012}.

We have estimated the absorbing column density assuming an absorbed
power-law model. Each source was classified as absorbed, or unabsorbed
at $1\sigma$ level, in the sense that those sources whose $1\sigma$
uncertainty interval is fully above $10^{22}$ cm$^{-2}$ or not
(respectively). And finally, we further explore possible
  mismatches in the observed X-ray absorption distributions for these
  subsamples (IR power-law and IR non-power-law) at three intrinsic
  rest-frame 2-10 keV luminosity bins ($<10^{43}$ erg/s, $10^{43-44}$
  erg/s, and $>10^{44}$ erg/s).  The goal was to investigate whether
the IR power-law criteria selects more absorbed AGN or not, and
specifically whether it is effective at selecting type-2 AGN at a
given luminosity range. The main results from our work are as follows:

\begin{itemize}
    \item All high-X-ray-spectral-quality sources in the deepest
      \xmm{} field have counterparts in the \emph{Spitzer}/IRAC bands,
      therefore using mid-IR data to search for obscured
      sources does not leave out any candidates.
    \item With our absorbed-power-law X-ray spectral analysis, 21 out
      of 147 sources are heavily absorbed
      ($N_H^{intr}>3\times10^{23}$cm$^{-2}$) but, when taking the
      uncertainties into account, we cannot confirm the Compton-Thick
      nature of any of our sources. We are, however, in full agreement
      with other papers using the same deep \xmm{} data, which use
      additional criteria to that end.
    \item At more than 3$\sigma$ level, we find that the fraction of
      absorbed sources among the IR power-law populations
      $(68^{-9}_{+8}\%)$ appears significantly higher than that for IR
      non-power-law galaxies $(48^{-10}_{+9}\%)$.
    \item We also found that the fraction of absorbed sources appears
      roughly constant $(\sim70)$ with luminosity for the IR power-law
      AGN, while it grows from $\sim20\%$ to $\sim65\%$ for the IR
      non-power-law galaxies with increasing luminosity.
    \item The main difference in the absorbed fraction between the IR
      power-law and the IR non-power-law sources happens at the lowest
      X-ray luminosities ($<10^{43}$ erg/s).  We understand this in
      terms of contrast with the host galaxy, in the sense that type-2
      AGN (in principle absorbed in X-rays) are more easily picked up
      by those criteria than type-1 AGN (unabsorbed) at low luminosities.
\end{itemize}

We conclude that the \citet{Donley2012} IR power-law criteria, if
admittedly incomplete, favour the selection of absorbed sources among the
X-ray detected AGN. This is particularly clear at low X-ray
luminosities. This prompts the question about the nature of the IR
power-law sources in the \xmm{} area without X-ray detection. Since we
do not miss any X-ray-detected sources by using mid-IR data and about 2/3
of the selected sources turn out to be absorbed, it is likely that the
mid-IR power-law criteria would pinpoint absorbed AGN among X-ray
undetected sources. These sources would still be detected in the mid-IR
(from the reprocessing of the AGN radiation), but they would have high
absorbing column densities, placing them beyond the current
capabilities of our most powerful observatories pushed to their limit.

\begin{acknowledgements}
  We are grateful to the referee for comments that helped improve
    the paper. This work is based on observations obtained with
  \xmm, an ESA science mission with instruments and contributions
  directly funded by ESA Member States and NASA. NC-M, FJC, SM and XB
  acknowledge financial support provided by the Spanish Ministry of
  Economy and Competitiveness through grant AYA2010-21490-C02-01. SM, 
  FJC and A.A.-H. acknowledge financial support by the Spanish Ministry 
  of Economy and Competitiveness through 
  grants AYA2010-21490-C02-01 and AYA2012-31447. SM acknowledges financial 
  support from the JAE-Doc program (Consejo Superior de Investigaciones 
  Cientficas, cofunded by FSE). AA-H
  acknowledges support from the Universidad de Cantabria through the
  Augusto G. Linares program. PGP-G acknowledges support from the
  Spanish Programa Nacional de Astronom\'{\i}a y Astrof\'{\i}sica
  under grants AYA2009-07723-E and AYA2009-10368. This work has made
  use of the Rainbow Cosmological Surveys Database, which is operated
  by the Universidad Complutense de Madrid (UCM).  We acknowledge
  financial contribution from the agreement ASI-INAF I/009/10/0 and
  from the INAF-PRIN-2011. \end{acknowledgements}

\bibliographystyle{aa}
\bibliography{References}

\end{document}